\input harvmac 
\input epsf.tex

\overfullrule=0mm
\def\IR{\relax{\rm I\kern-.18em R}} 
\font\cmss=cmss10
\font\cmsss=cmss10 at 7pt 
\def\IZ{\relax\ifmmode\mathchoice
{\hbox{\cmss Z\kern-.4em Z}}{\hbox{\cmss Z\kern-.4em Z}}
{\lower.9pt\hbox{\cmsss Z\kern-.4em Z}} 
{\lower1.2pt\hbox{\cmsss Z\kern-.4em Z}}
\else{\cmss Z\kern-.4em Z}\fi}
\def\file#1{#1}
\def\figbox#1#2{\epsfxsize=#1\vcenter{
\epsfbox{\file{#2}}}} 
\newcount\figno
\figno=0
\def\fig#1#2#3{
\par\begingroup\parindent=0pt\leftskip=1cm\rightskip=1cm\parindent=0pt
\baselineskip=11pt
\global\advance\figno by 1
\midinsert
\epsfxsize=#3
\centerline{\epsfbox{#2}}
\vskip 12pt
{\bf Fig. \the\figno:} #1\par
\endinsert\endgroup\par
}
\def\figlabel#1{\xdef#1{\the\figno}}
\def\encadremath#1{\vbox{\hrule\hbox{\vrule\kern8pt\vbox{\kern8pt
\hbox{$\displaystyle #1$}\kern8pt}
\kern8pt\vrule}\hrule}}

\def\tvi{\vrule height 10pt depth 6pt width 0pt}
\def\tv{\tvi\vrule}

\font\cmss=cmss10 \font\cmsss=cmss10 at 7pt
\def\IZ{\relax\ifmmode\mathchoice
{\hbox{\cmss Z\kern-.4em Z}}{\hbox{\cmss Z\kern-.4em Z}}
{\lower.9pt\hbox{\cmsss Z\kern-.4em Z}}
{\lower1.2pt\hbox{\cmsss Z\kern-.4em Z}}\else{\cmss Z\kern-.4em Z}\fi}

\Title{\vbox{\hsize=3.truecm \hbox{SPhT/99-131}}}
{{\vbox {
\bigskip
\centerline{Matrix Model Combinatorics:}
\medskip
\centerline{Applications to Folding and Coloring}
}}}
\bigskip
\centerline{P. Di Francesco\foot{philippe@spht.saclay.cea.fr}}
\medskip
\centerline{ \it CEA-Saclay, Service de Physique Th\'eorique,}
\centerline{ \it F-91191 Gif sur Yvette Cedex, France}
\medskip

\vskip .5in


We present a detailed study of the combinatorial interpretation of
matrix integrals, including the examples of tessellations of arbitrary genera,
and loop models on random surfaces. After reviewing their methods of solution, we 
apply these to the study of various folding problems arising from physics, 
including: the meander (or polymer folding)
problem ``enumeration of all topologically inequivalent closed non-intersecting 
plane curves intersecting a line through a given number of points" and
a fluid membrane folding problem reformulated as that of ``enumerating  
all vertex-tricolored triangulations of arbitrary genus, with given numbers
of vertices of either color".


\noindent
\Date{11/99}
\noindent {1.} {Introduction} \leaderfill{2} \par 
\noindent {2.} {Matrix Models and Combinatorics} \leaderfill{4} \par 
\noindent \quad{2.1.} {Gaussian Integrals and Wick's Theorem} \leaderfill{4} \par 
\noindent \quad{2.2.} {One-Hermitian Matrix Model: Discrete Random Surfaces} \leaderfill{6} \par 
\noindent \quad{2.3.} {Multi-Hermitian Matrix Case} \leaderfill{12} \par 
\noindent \quad{2.4.} {A generating Function for Fatgraphs} \leaderfill{16} \par 
\noindent {3.} {Matrix Models: Solutions} \leaderfill{18} \par 
\noindent \quad{3.1.} {Reduction to Eigenvalues} \leaderfill{18} \par 
\noindent \quad{3.2.} {Orthogonal Polynomials} \leaderfill{19} \par 
\noindent \quad{3.3.} {Large $N$ asymptotics I: Orthogonal Polynomials} \leaderfill{21} \par 
\noindent \quad{3.4.} {large $N$ asymptotics II: Saddle-Point Approximation} \leaderfill{23} \par 
\noindent \quad{3.5.} {Critical Behavior and Asymptotic Enumeration} \leaderfill{29} \par 
\noindent \quad{3.6.} {Gaussian Words} \leaderfill{30} \par 
\noindent {4.} {Folding Polymers: Meanders} \leaderfill{32} \par 
\noindent \quad{4.1.} {Definitions and Generalities} \leaderfill{32} \par 
\noindent \quad{4.2.} {A Simple Algorithm; Numerical Results} \leaderfill{36} \par 
\noindent {5.} {Algebraic Formulation: Temperley-Lieb Algebra} \leaderfill{38} \par 
\noindent \quad{5.1.} {Definition} \leaderfill{38} \par 
\noindent \quad{5.2.} {Meander Polynomials} \leaderfill{40} \par 
\noindent \quad{5.3.} {Meander Determinants} \leaderfill{41} \par 
\noindent {6.} {Matrix Model for Meanders} \leaderfill{41} \par 
\noindent \quad{6.1.} {The B$\&$W Model} \leaderfill{42} \par 
\noindent \quad{6.2.} {Meander Polynomials and Gaussian Words} \leaderfill{44} \par 
\noindent \quad{6.3.} {Exact Asymptotics for the case of arbitrary many rivers} \leaderfill{47} \par 
\noindent \quad{6.4.} {Exact Meander Asymptotics from Fully-Packed Loop Models coupled to Two-dimensional Quantum Gravity} \leaderfill{50} \par 
\noindent {7.} {Folding Triangulations} \leaderfill{54} \par 
\noindent \quad{7.1.} {Folding the Triangular Lattice} \leaderfill{55} \par 
\noindent \quad{7.2.} {Foldable Triangulations} \leaderfill{56} \par 
\noindent {8.} {Exact Solution} \leaderfill{59} \par 
\noindent \quad{8.1.} {Discrete Hirota Equation} \leaderfill{59} \par 
\noindent \quad{8.2.} {Direct Expansion and Large N Asymptotics} \leaderfill{63} \par 
\noindent {9.} {Conclusion} \leaderfill{64} \par 

\writetoc

\nref\NOB{G. 't Hooft, Nucl. Phys. {\bf B72} (1974) 461}
\nref\SAPO{E. Br\'ezin, C. Itzykson, G. Parisi and J.-B. Zuber, Comm. Math. 
Phys. {\bf 59} (1978) 35.}
\nref\GRAVI{see for instance the review  by P. Di Francesco, 
P. Ginsparg and J. Zinn-Justin,
{\it 2D Gravity and Random Matrices}, 
Physics Reports {\bf 254} (1995) 1-131, and references therein.}
\nref\KPZ{V.G. Knizhnik, A.M. Polyakov and A.B. Zamolodchikov, Mod. Phys. Lett.
{\bf A3} (1988) 819; F. David, Mod. Phys. Lett. {\bf A3} (1988) 1651; J.
Distler and H. Kawai, Nucl. Phys. {\bf B321} (1989) 509.}
\nref\BERT{B. Duplantier, {\it Two-dimensional copolymers and 
exact conformal multifractality}, preprint cond-mat/9812439;
{\it Exact multifractal exponents for two-dimensional
percolation}, preprint cond-mat/9901008; 
{\it Conformally invariant fractals and
potential theory}, preprint cond-mat/9908314.}
\nref\ORTHO{D. Bessis, Comm. Math. Phys. {\bf 69} (1979) 147-163;
D. Bessis, C.Itzykson and J.-B. Zuber, Adv. in Appl.
Math. {\bf 1} (1980) 109.}
\nref\VARFO{
A. Sainte-Lagu\"e,
{\it Avec des nombres et des lignes (R\'ecr\'eations Math\'ematiques)},
Vuibert, Paris (1937);
J. Touchard, {\it Contributions \`a l'\'etude du probl\`eme
des timbres poste}, Canad. J. Math. {\bf 2} (1950) 385-398;
W. Lunnon, {\it A map--folding problem},
Math. of Computation {\bf 22}
(1968) 193-199;
K. Hoffman, K. Mehlhorn, P. Rosenstiehl and R. Tarjan, {\it
Sorting Jordan sequences in linear time using level-linked search
trees}, Information and Control {\bf 68} (1986) 170-184;
V. Arnold, {\it The branched covering of $CP_2 \to S_4$,
hyperbolicity and projective topology},
Siberian Math. Jour. {\bf 29} (1988) 717-726;
K.H. Ko, L. Smolinsky, {\it A combinatorial matrix in
$3$-manifold theory}, Pacific. J. Math {\bf 149} (1991) 319-336.}
\nref\LZ{S. Lando and A. Zvonkin, {\it Plane and Projective Meanders},
Theor. Comp.  Science {\bf 117} (1993) 227-241, and {\it Meanders},
Selecta Math. Sov. {\bf 11} (1992) 117-144.}
\nref\BACH{R. Bacher, {\it Meander Algebras}, pr\'epublication de l'Institu Fourier
n$^o$ $478$ (1999).}
\nref\DGG{P. Di Francesco, O. Golinelli and E. Guitter, {\it Meander,
folding and arch statistics}, Mathl. Comput. Modelling {\bf 26} (1997) 97-147.}
\nref\NOUS{P. Di Francesco, O. Golinelli and E. Guitter, {\it Meanders:
a direct enumeration approach}, Nuc. Phys. {\bf B 482} [FS] (1996) 497-535.}
\nref\NTLA{P. Di Francesco, O. Golinelli and E. Guitter, {\it
Meanders and the Temperley-Lieb algebra}, Commun.Math.Phys. {\bf 186} (1997) 1-59.}
\nref\MEDET{P.\ Di Francesco, {\it SU(N) Meander Determinants},
J.\ Math.\ Phys.\ 38 (1997) 5905-5943, and {\it Meander Determinants},
Commun.\ Math.\ Phys.\ 191 (1998) 543-583.}
\nref\TRUNC{P. Di Francesco, {\it Truncated Meanders},
preprint UNC-CH-MATH-98/3, to appear in
proceedings of the AMS (1999).}
\nref\MAK{Y. Makeenko, {\it Strings, Matrix Models and Meanders}, proceedings
of the 29th Inter. Ahrenshoop Symp., Germany (1995); 
Y. Makeenko and H. Win Pe, {\it Supersymmetric matrix models and the meander
problem}, preprint ITEP-TH-13$/$95 (1996); G. Semenoff and
R. Szabo {\it Fermionic Matrix Models} preprint UBC$/S96/2$ (1996).}
\nref\Goli{O. Golinelli, {\it A Monte-Carlo study of meanders}, preprint
cond-mat/9906329, to appear in EPJ {\bf B} (2000). }
\nref\JEN{I. Jensen, {\it Enumerations of Plane Meanders}, preprint
cond-mat/9910313.}
\nref\ASY{P. Di Francesco, O. .Golinelli and E. Guitter, {\it Meanders:
Exact Asymptotics}, preprint cond-mat/9910453.}
\nref\ON{ I. Kostov, Mod. Phys. Lett. {\bf A4} (1989) 217; M. Gaudin and
I. Kostov, Phys. Lett. {\bf B220} (1989) 200; I. Kostov and M. Staudacher, Nucl.
Phys. {\bf B384} (1992) 459.}
\nref\QPO{ V. Kazakov, Nucl. Phys. {\bf B4} (Proc. Suppl.) (1998), 93; J.-M. Daul,
preprint hep-th/9502014; P. Zinn-Justin, preprint cond-mat/9903385; 
B. Eynard and G. Bonnet, {\it The Potts-q random matrix model:
loop equations, critical exponents, and rational case}, preprint cond-mat/9906130.}
\nref\FAT{P. Di Francesco and C. Itzykson, {\it A Generating Function for Fatgraphs},
Annales de l'Institut Henri Poincar\'e, Vol. {\bf 59}, no. 2 (1993) 117-139;
V. Kazakov, M. Staudacher and T. Wynter, {\it Character expansion methods
and for matrix models of dually weighted graphs}, Comm. Math. Phys. {\bf 177} (1996)
451-468, {\it Almost flat planar diagrams}, Comm. Math. Phys. {\bf 179} (1996) 235-256
and {Exact solution of discrete two-dimensional $R^2$ gravity}, Nucl. Phys. {\bf B471} (1996)
309-333.}
\nref\ONSOL{B. Eynard and J. Zinn-Justin, Nucl. Phys. {\bf B386} (1992) 558;
B. Eynard and C. Kristjansen, Nucl. Phys. {\bf B455} (1995) 577 and Nucl. Phys. 
{\bf B466} (1996) 463-487.}
\nref\CFT{ P. Di Francesco, P. Mathieu and D. S\'en\'echal,
{\it Conformal Field Theory}, Graduate Texts in Contemporary Physics,
Springer (1996).}
\nref\NIN{B.\ Nienhuis in {\it Phase Transitions and Critical Phenomena},
Vol.\ 11, eds.\ C.\ Domb and J.L.\ Lebowitz, Academic Press 1987.}
\nref\CK{L. Chekhov and C. Kristjansen, {\it Hermitian Matrix Model with Plaquette
Interaction}, Nucl.Phys. {\bf B479} (1996) 683-696.}
\nref\JACO{J. Jacobsen and J. Kondev, {\it Field theory of compact polymers
on the square lattice}, Nucl. Phys. {\bf B 532} [FS], (1998) 635-688,
{\it Transition from the compact to the dense phase of two-dimensional polymers},
J. Stat. Phys. {\bf 96}, (1999) 21-48.}
\nref\DUD{F. David and B. Duplantier, {\it Exact partition functions and correlation
functions of multiple Hamiltonian walks on the Manhattan lattice}, J. Stat.
Phys. {\bf 51}, (1988) 327-434.}
\nref\DGTRI{P.\ Di Francesco and E.\ Guitter, Europhys.\ Lett.\ 26
(1994) 455.}
\nref\TRICO{P.\ Di\ Francesco, B.\ Eynard and E.\ Guitter, 
{\it Coloring Random Triangulations},
Nucl. Phys. {\bf B516 [FS]} (1998) 543-587.}
\nref\BAX{R.J. Baxter, J. Math. Phys. {\bf 11} (1970) 784 and
J. Phys. {\bf A19} Math. Gen. (1986) 2821.}
\nref\SQDI{P. Di Francesco, {\it Folding the Square-Diagonal Lattice},
Nucl. Phys. {\bf B525[FS]} (1998) 507-548; {\it Folding Transitions of the Square-Diagonal
Lattice}, Nucl. Phys. {\bf B528[FS]} (1998) 453-465;
P. Di Francesco and E. Guitter, 
{\it Folding Transition of the Triangular Lattice},
Phys. Rev. {\bf E50} (1994) 4418; M. Bowick, P. Di Francesco,
O. Golinelli and E. Guitter,  {\it 3D Folding of the triangular lattice}, 
Nucl. Phys. {\bf B450[FS]} (1995) 463-494.}
\nref\AMS{P. Di Francesco, {\it Folding and Coloring Problems in Mathematics 
and Physics}, preprint UNC-CH-MATH-98/4 (1998).}
\nref\MIT{Mathematical Intelligencer Volume 19 Number 4 (1997) 48;
Volume 20 Number 3 (1998) 29.}
\nref\WIG{I. Krichever, O. Lipan, P. Wiegmann and A. Zabrodin,
{\it Quantum integrable systems and elliptic solutions of classical
discrete nonlinear equations},
Comm. Math. Phys. {\bf 188} (1997) 267.}
\nref\ITZ{C. Itzykson and J.-B. Zuber, {\it The planar approximation II},
J. Math. Phys. {\bf 21} (1980) 411.}
\nref\DH{Harish-Chandra, {\it Differential operators on a semi-simple
Lie algebra}, Amer. Jour. of Math {\bf 79} (1957) 87;
J. Duistermaat and G. Heckman, {\it On the variation of cohomology of
the symplectic form of the reduced phase space},
Inv. Math. {\bf 69} (1982) 259-268.}
\nref\TUT{W. Tutte, {\it A Census of Planar Maps},
Canad. Jour. of Math. {\bf 15} (1963) 249.}


\newsec{Introduction}

Our first aim in these notes is to convince the reader that matrix integrals, 
exactly calculable or not,
can always be interpreted in some sort of combinatorial way as generating functions for 
decorated graphs of given genus, with possibly specified vertex and/or face valencies. 
We show this by expressing pictorially the processes involved in computing 
Gaussian integrals over matrices, what physicists call generically Feynman rules.
These matrix diagrammatic techniques have been first developed in the context of 
quantum chromodynamics in the limit of large number of colors (the size of the matrix)
\NOB\ \SAPO,
and more recently in the context of two-dimensional quantum gravity, namely the coupling of
two-dimensional statistical models (matter theories) to the fluctuations of
the two-dimensional space into surfaces of arbitrary topologies (gravity) \GRAVI. 
These toy models for non-critical string theory are a nice testing ground for 
physical ideas, and have led to many confirmations of continuum field-theoretical
results in quantum gravity. The purely combinatorial aspect of these models has
often been treated as side-result, and we believe it deserves more attention,
especially in view of some spectacular results. Indeed, the non-critical string
machinery allows one to relate critical properties (such as singularities of 
thermodynamic quantities) of the flat space statistical models to those of the
same models defined on random surfaces \KPZ. This has led for instance to recent progress 
in the study of random walks, by using the inverse relation to deduce flat
space results from gravitational ones \BERT.   

Once the connection is made between a combinatorial graph-related problem and
some matrix model, we still have to compute the integral. Various powerful techniques
have been developed, mainly in the context of various branches of physics, to compute
those integrals: the original one is orthogonal polynomials \ORTHO, but only applies to 
``simple" models. More general is the saddle-point technique \SAPO, that however only
allows for computing these integrals in the limit of large size of the matrices. 

In these notes, we wish to present applications of the combinatorics of 
matrix models to some specific questions arising in physics having to do with
folding.
Folding problems arise in biology and physics in particular when considering 
polymers or membranes. An ideal polymer is a chain of say $n$ identical constituents
represented by segments (chemical bonds) attached to one another by their ends
(atoms), around which
they can rotate freely. Membranes are two-dimensional generalizations of polymers,
i.e. reticulated networks made of vertices (atoms) linked by edges (chemical bonds), either in
the form of a regular lattice (tethered membranes) or in the form of networks with 
arbitrary vertex valencies (fluid membranes).
Ideally, imposing that all bonds be rigid, the only possibility for a polymer
or membrane to change its spatial configuration is through folding, in which 
atoms (for polymers) or bonds (for membranes) serve as hinges. Quantities of interest for
physicists are thermodynamic ones, characteristic of the systems when their size is
large.  
Mathematically, these correspond to asymptotics of say the numbers of distinct 
configurations of folding of polymers or membranes of given length or area,
when the latter tend to infinity.
Both in the case of polymer and fluid membrane folding, we will present 
matrix models allowing for the calculation of such thermodynamic properties.

The polymer folding problem is better known in mathematics as the ``meander problem",
namely that of enumerating all the topologically inequivalent configurations of
a closed nonselfintersecting plane curve intersecting a line through a given number
of points. This problem apparently first emerged in some work by Poincar\'e
in the beginning of the century, and reemerged in various areas of
mathematics \VARFO\ \LZ\ \BACH, from recreational mathematics
to the 16th Hilbert problem to computer science to 
the theory of knots and links. 
In physics, the formulation as a folding problem and the relation to matrix
integrals \DGG\ have brought very different developments both algebraic \NTLA\ \MEDET\ \TRUNC\
\MAK\  and numerical \NOUS\ \Goli\ \JEN.
As a highly non-trivial outcome of our study, we will present the exact
meander asymptotics recently derived in \ASY.

The notes are organized as follows. 
In Part A, we give a detailed presentation of the combinatorial interpretation
of Hermitian matrix integrals, as generating tools for fatgraphs (Sect. 2),
and proceed to review their various methods of solution (Sect. 3).
Part B is devoted to the application of this fact to the problem of 
enumeration of all distinct compact folding configurations of a 
closed or open polymer, namely the number of distinct ways to fold a 
(self-avoiding) chain of identical
constituents onto itself: this is also known as the meander or semi-meander problem. 
After defining the problem and reviewing a few known
results (Sect. 4), we first present an algebraic formulation of the counting problem
within the
framework of the Temperley-Lieb algebra, omnipresent in the integrable statistical
models, as well as the theory of knots and links (Sect. 5). This connection 
produces remarkable results, like the exact expression for the meander determinant,
a meander-related quantity.     
In Sect. 6, we connect the meander and related problems to a multimatrix integral,
that allows for the exact determination of meandric configuration exponents, 
in particular through the abovementioned connection \KPZ\ beetween flat and curved space
models.
Finally, we study in Part C a matrix model for generating foldable triangulations,
a sort of two-dimensional generalization of the meander problem, but without the
self-avoidance constraint. These triangulations form a simple model for fluid membranes,
encountered in physics and biology.
After posing the problem (Sect. 7), we introduce a two-matrix integral generating
these triangulations, that we solve in various ways (Sect. 8). Finally,
Sect. 9 gathers a few
concluding remarks.

\vskip 1.truecm 

\centerline{\bf PART A: Matrix Model Combinatorics}
\vskip 1.truecm

The aim of this introductory part is to familiarize the reader with 
the use of matrix integrals as tools for generating and enumerating 
graphs with various decorations. These correspond in turn to physical
models of matter coupled to two-dimensional quantum gravity, in the
form of fluctuating surfaces of arbitrary genus.  
In the following, we first present the combinatorial tools and give
recipees to construct ad-hoc matrix integrals for various graph enumeration
problems. We then expose various methods of computation of these 
matrix integrals, concentrating in particular on the planar graph limit.

\newsec{Matrix Models and Combinatorics} 

\subsec{Gaussian Integrals and Wick's Theorem}

Consider the following Gaussian average
\eqn\gauint{ \langle x^{2n} \rangle = {1\over \sqrt{2 \pi}}
\int_{-\infty}^{\infty} e^{-{x^2 \over 2}}  x^{2n} dx  
=(2n-1)!!={(2n)!\over 2^n n!} }
Among the many ways to compute this integral, let us choose the
so-called source integral method, namely define the Gaussian source
integral
\eqn\source{\Sigma(s)=\langle e^{xs}\rangle={1\over \sqrt{2\pi}} 
\int_{-\infty}^{\infty} e^{-{x^2\over 2}+sx} dx= e^{{s^2\over 2}}}
Then the average \gauint\ is obtained by taking $2n$ derivatives
of $\Sigma(s)=e^{{s^2\over 2}}$ wrt $s$ and by setting $s=0$ in the end. 
It is then immediate
to see that these derivatives must be taken by {\it pairs}, in which
one derivative acts on the exponential and the other one on the prefactor
$s$. Parallelly, we note that $(2n-1)!!=(2n-1)(2n-3)...3.1$ is the 
total number of distinct associations of $2n$ objects
into $n$ pairs. 
We may therefore formulate pictorially the computation of \gauint\ as 
follows. 
\fig{A star-diagram with one vertex and $2n$ out-coming
half-edges stands for the integrand $x^{2n}$. In the second
diagram,
we have represented one non-zero contribution to $\langle
x^{2n}\rangle$ obtained by taking derivatives of $\Sigma(s)$ by
pairs represented as the corresponding connections of half-edges
(the
$x$'s and $s$'s are dual to one another, hence derivatives wrt $s$
are in one to one correspondence with insertions of $x$ in the
integrand).}{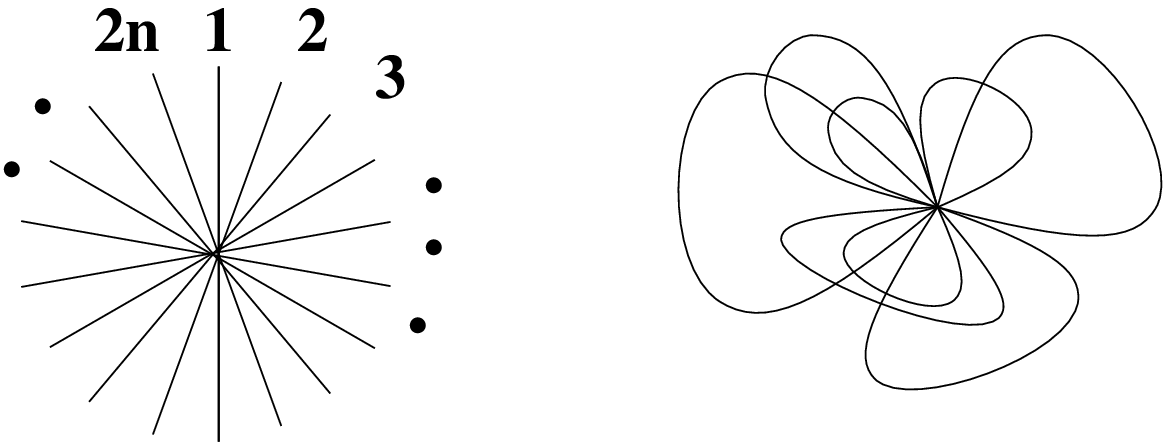}{6.cm}
\figlabel\diagone
We first draw a star-graph (see Fig.\diagone), with one central vertex and 
$2n$ outcoming
half-edges labelled $1$ to $2n$ clockwise, one for each $x$ in the integrand 
(this amounts to labelling the $x$'s in $x^{2n}$
from $1$ to $2n$). Now the pairs of derivatives
taken on the source integral are in one-to-one correspondence with pairs
of edges in the pictorial representation. Moreover, to get a non-zero 
contribution to
$\langle x^{2n}\rangle$, we must saturate the set of $2n$ legs by
taking $n$ pairs of them. Let us represent each such saturation
by connecting the corresponding edges as in Fig.\diagone. 
We get exactly $(2n-1)!!$
distinct closed star-graphs with one vertex. We may therefore write the
one-dimensional version of Wick's theorem
\eqn\wione{ \langle x^{2n} \rangle = \sum_{{\rm pairings}} \prod 
\langle x^2 \rangle }
where the sum extends over all pairings saturating the $2n$ half-edges,
and the weight is simply the product over all the edges formed
of the corresponding averages $\langle x^2\rangle =(d^2/ds^2)
\Sigma(s)\vert_{s=0}=1$.
Each saturation forms a ``Feynman diagram" of the Gaussian average. The 
edge pairings are called propagators (with value $1$ here). The
Feynman rules are simply the set of values of these propagators.
This may appear like a complicated way of writing a rather trivial result,
but it suits our purposes for generalization to matrix models and graphs.
Let us mention that the pictorial interpretation we have given
for the computation of $\langle x^{2n}\rangle$ is not unique. For instance,
we could have arbitrarily split the average into $\langle x^p x^q\rangle$
for some integrers $p,q$ with $p+q=2n$. Then we would have rather represented
{\it two} vertices with respectively $p$ and $q$ out-coming half-edges.
Wick's theorem \wione\ states that we must now sum over (possibly disconnected)
graphs obtained by saturating the $p+q$ half-edges by pairs. The result of
course is still the same.

As a last remark, and to make the contact with graphs,
we may consider for instance the formal series expansion
\eqn\forser{ z(g_1,g_2,...)=\langle e^{\sum_{i\geq 1} g_i{x^i}} \rangle }
in powers of $g_1,g_2,...$,
whose coefficients are computed using \gauint. Thanks to the previous
pictorial interpretation, we may compute a typical term in the series
expansion, say the coefficient $\langle \prod (x^i)^{V_i}\rangle$
of $\prod g_i^{V_i}/V_i!$, by drawing 
$V_i$ star-diagrams with $i$ half-edges, $i=1,2,...$, and saturating
the $2E=\sum iV_i$ half-edges in all possible ways by forming
$E$ pairs. Hence computing \forser\ is reinterpreted as a generating function of 
graphs with specified numbers of vertices of given valencies.

\subsec{One-Hermitian Matrix Model: Discrete Random Surfaces}

Let us now repeat the calculations of the previous section with 
the following Gaussian Hermitian matrix average of an arbitrary
function $f$
\eqn\gaumat{ \langle f(M) \rangle = {1\over Z_0(N)} 
\int dM e^{-N Tr {M^2\over 2}}
f(M) }
where the integral extends over Hermitian $N\times N$ matrices, with the 
standard Haar measure $dM=\prod_i dM_{ii} \prod_{i<j} dRe(M_{ij}) dIm(M_{ij})$,
and the normalization factor $Z_0(N)$ is fixed by requiring that 
$\langle 1\rangle=1$ for $f=1$. Typically, we may take for $f$ a monomial
of the form $f(M)=\prod_{(i,j)\in I} M_{ij}$, $I$ a finite set of pairs
of indices. Note the presence of the normalization factor $N$ (=the size
of the matrices) in the exponential. Note also the slight abuse of notation
as we still denote averages with the same bracket sign as in previous section:
we may simply include the case of the previous section as the particular case
of integration over $1\times 1$ Hermitian matrices (i.e. real numbers) here.

Like in the one-dimensional case of the previous section, for 
a given Hermitian $N\times N$ matrix $S$, let us
introduce the source integral 
\eqn\sourmat{ \Sigma(S)=\langle e^{Tr(SM)} \rangle = e^{{Tr(S^2)\over 2N}} }
easily obtained by completing the square $M^2 -N(SM+MS)=(M-NS)^2-N^2 S^2$
and performing the change of variable $M'=M-NS$.
We can use \sourmat\ to compute any average of the form
\eqn\avform{ \langle M_{ij} M_{kl} ... \rangle={\partial \over
\partial S_{ji}} {\partial \over \partial S_{lk}} ... \ \Sigma(S)\big\vert_{S=0}}
Note the interchange of the indices due to the trace $Tr(MS)=\sum M_{ij}S_{ji}$.
As before, derivatives wrt elements of $S$ must go by pairs, one
of which acts on the exponential and the other one on the $S$ element thus 
created. In particular, a fact also obvious from the parity of the Gaussian,
\avform\ vanishes unless there are an even number of matrix elements of $M$
in the average. In the simplest case of two matrix elements, we have
\eqn\prop{ \langle M_{ij} M_{kl}\rangle = {\partial\over
\partial S_{lk}} {1\over N} S_{ij} e^{{Tr(S^2)\over 2N}}\bigg\vert_{S=0}=
{1\over N}\delta_{il}\delta_{jk} }
Hence the pairs of derivatives must be taken with respect to $S_{ij}$ and
$S_{ji}$ for some pair $i,j$ of indices to yield a non-zero result. 
This leads naturally to the Matrix Wick's theorem:
\eqn\wimat{ \langle \prod_{(i,j)\in I} M_{ij} \rangle=
\sum_{\rm pairings\ P} \prod_{
(ij), (kl) \in P} \langle M_{ij} M_{kl}\rangle }
where the sum extends over all pairings saturating the (pairs of) indices
of $M$ by pairs. 

We see that in general, due to the restrictions 
\prop\ many terms in \wimat\ will vanish. Let us now give a 
pictorial interpretation for the non-vanishing contributions in \wimat.
We represent a matrix element $M_{ij}$ as a half-edge (with a marked end)
made of a double-line, each of which is oriented in an opposite direction.
We decide that the line pointing from the mark carries the index $i$,
while the other one, pointing to the mark, carries the index $j$.
This reads
\eqn\markedge{ M_{ij}\ \  \leftrightarrow \ \ \figbox{2.cm}{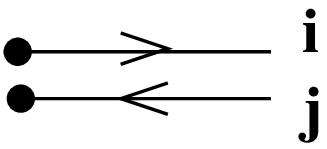} }
The two-element result \prop\ becomes simply the construction of an
edge (with both ends marked) out of two half-edges $M_{ij}$ and $M_{kl}$, 
but is non-zero 
only if the indices $i$ and $j$ are conserved along the oriented lines.
This gives pictorially
\eqn\propag{ \langle M_{ij} M_{ji} \rangle\ \   \leftrightarrow \ \ 
\figbox{3.5cm}{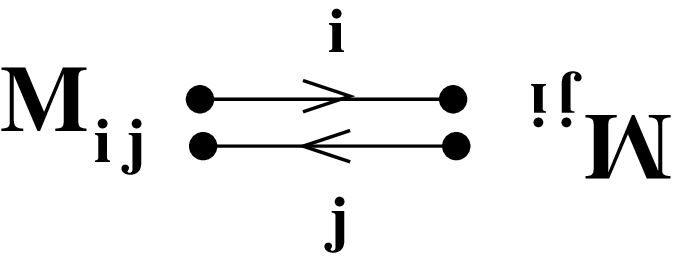} }
Similarly, an expression of the form Tr$(M^n)$ will be represented as
a star-diagram with one vertex connected to $n$ double half-edges
in such a way as to respect the identification of the various running indices,
namely
\eqn\tracrep{ {\rm Tr}(M^n)=\sum_{i_1,i_2,...,i_n} M_{i_1i_2}M_{i_2i_3}...
M_{i_ni_1} \leftrightarrow \figbox{2.5cm}{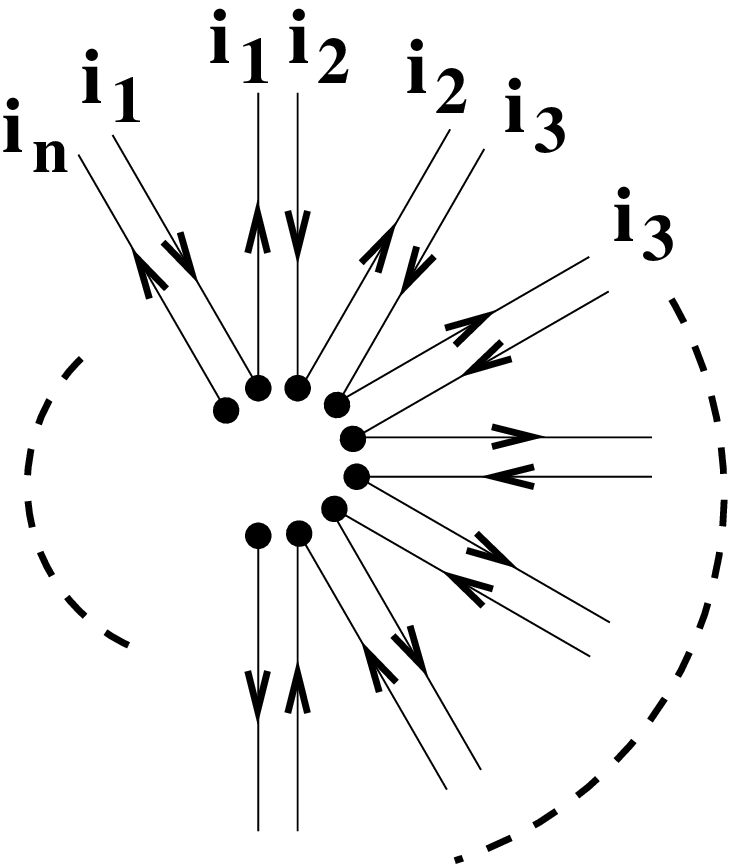} }

\fig{An example of planar (petal) diagram (a) and a non-planar one
(b). Both diagrams have $n=2p=12$ half-edges, connected with $p=6$
edges. The diagram (a) has $p+1=7$ faces bordered by oriented loops,
whereas (b) only has $3$ of them. The Euler characteristic
reads $2-2h=F-E+1$ ($V=1$ in both cases), and gives the genus $h=0$ for (a),
and $h=2$ for (b).}{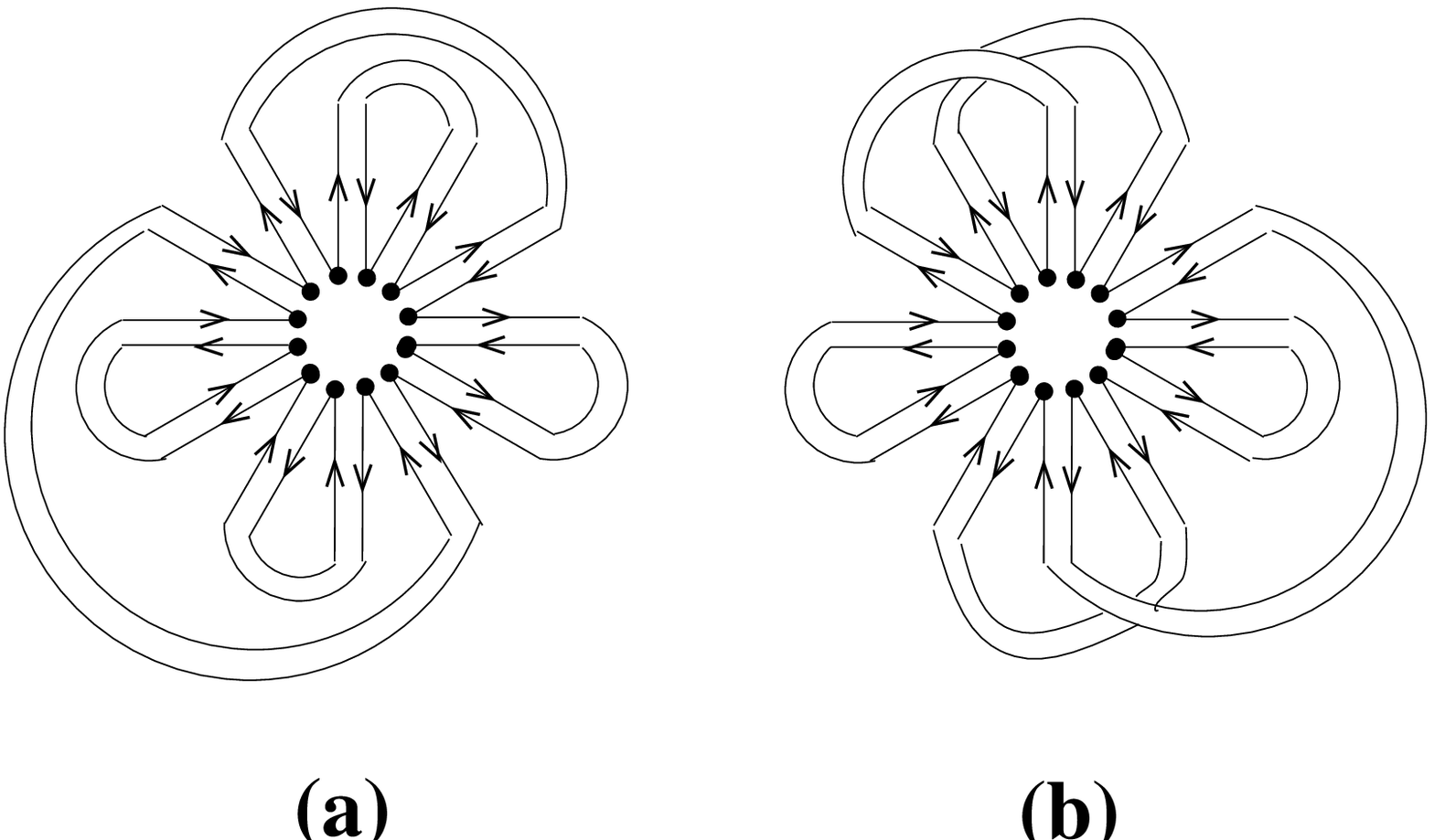}{7.cm}
\figlabel\planar

As a first application of this diagrammatic interpretation of the Wick
theorem \wimat, let us compute the large $N$ asymptotics of $\langle {\rm
Tr}(M^n)\rangle$. To compute $\langle {\rm
Tr}(M^n)\rangle$, we must first draw a star-diagram as in \tracrep, then
apply \wimat\ to express the result as a sum over the saturations of
the star with edges connecting its outcoming half-edges by pairs. 
To get a non-zero result, we must clearly have $n$ even, say $n=2p$.
Note then that
there are $(2p-1)!!$ such pairings, allowing us to recover the result of
the previous section by setting $N=1$, and simply replacing all oriented
double lines by unoriented single ones. But if instead we take $N$ to be 
large, we see that only a fraction of these $(2p-1)!!$ pairings will contribute
at leading order. Indeed, assume first we restrict the set of pairings to
``planar" ones (see Fig.\planar\ (a)), namely such that the saturated star diagrams
have a petal structure in which edges only connect pairs of half-edges of the
form $(ij)$, $i<j$ and $(kl)$, $k<l$ with either $j<k$ or $k<i<j<l$ or
$i<k<l<j$ (in the labelling of half-edges, we have taken cyclic boundary
conditions, namely the labels $1$ and $n+1$ are identified). In other words,
the petals are  either juxtaposed or included into one-another. 
We may compute the genus of the petal diagrams by noting that they form
a tessellation of the sphere (=plane plus point at infinity). This tessellation
has $V=1$ vertex (the star), $E=p$ edges, and $F$ faces, including the
``external" face containing the point at infinity. The planarity of the diagram
simply expresses that its genus $h$ vanishes, namely
\eqn\plapro{ 2-2h=2=F-E+V=F+1-p \ \ \Rightarrow \ \ F=p+1}
Such diagrams
receive a total contribution $1/N^{p}$ from the propagators (weight $1/N$ 
per connecting edge), but we still have to sum over the remaining matrix
indices $j_1,j_2,...,j_{p+1}$ running over the $p+1$ oriented loops we have
created, which form the boundaries of the
$F=p+1$ faces.  This gives an extra weight $N$ per
loop (or face) of the diagram, hence a total contribution of $N^{p+1}$. 
So all the petal diagrams
contribute the same total factor $N$ to $\langle {\rm Tr}(M^n)\rangle$. Now
any non-petal (i.e. non-planar, see Fig.\planar\ (b)) 
diagram must have at least {\it two less}
oriented loops, hence contributes at most $1/N$ to $\langle {\rm
Tr}(M^n)\rangle$. Indeed, its Euler characteristic is negative or zero, hence
it has $F\leq E-V=p-1$ and it contributes at most for $N^{F-p}\leq 1/N$.
So, to leading order in $N$, only the genus zero (petal) diagrams contribute. 
We simply have to count them. This is a standard problem in combinatorics:
one may for instance derive a recursion relation for the number $c_p$ of petal
diagrams with $2p$ half-edges, by fixing the left end of an edge (say at
position $1$), and summing over the positions of its right end (at positions
$2j$, $j=1,2,...,p$), and noting that the petal thus formed may contain
$c_{j-1}$ distinct petal diagrams and be next to $c_{p-j}$ distinct ones.
This gives the recursion relation
\eqn\recucat{ c_p=\sum_{j=1}^p c_{j-1} c_{p-j} \qquad c_0=1}
solved by the Catalan numbers
\eqn\catalan{ c_p={(2p)! \over (p+1)! p!} }
Finally, we get the one-matrix planar Gaussian average
\eqn\magau{ \lim_{N\to \infty} {1\over N} \langle {\rm Tr}(M^n) \rangle =
\left\{ \matrix{ c_p & {\rm if} \ \ n=2p \cr
0 & {\rm otherwise} \cr} \right. }
This exercise shows us what we have gained by considering $N\times N$ matrices
rather than numbers: we have now a way of discriminating between the various
genera of the graphs contributing to Gaussian averages. This fact will be
fully  exploited in the next example.

We would now like to present an application of \wimat\ with important physical
and mathematical consequences.
Let us apply the matrix Wick theorem \wimat\ to the following generating
function $f(M)=\exp(N\sum_{i\geq 1} g_i {\rm Tr}(M^i)/i)$, to be understood
as a formal power series of the $g_i$, $i=1,2,3,4,...$
\eqn\gram{\eqalign{ Z_N(g_1,g_2,...)&=\langle e^{N\sum_{i\geq 1} 
g_i {\rm Tr}({M^i\over i})} \rangle\cr
&=\sum_{n_1,n_2,...\geq 0} \prod_{i\geq 1} {(Ng_i)^{n_i}\over i^{n_i} n_i!} \langle
\prod_{i\geq 1} {\rm Tr}(M^i)^{n_i} \rangle\cr
&=\sum_{n_1,n_2,...\geq 0}\prod_{i\geq 1} {(Ng_i)^{n_i}\over i^{n_i} n_i!} 
\sum_{{\rm all}\ {\rm labelled}\ {\rm fatgraphs}\ \Gamma \atop
{\rm with} \ n_i\ i-{\rm valent}\ {\rm vertices}} N^{-E(\Gamma)} N^{F(\Gamma)}\cr} }
by direct application \wimat. 
\fig{A typical connected fatgraph $\Gamma$, corresponding to the 
average $\langle {\rm Tr}(M)^3
{\rm Tr}(M^2)^2{\rm Tr}(M^3)
{\rm Tr}(M^4)^2 {\rm Tr}(M^6){\rm Tr}(M^8)\rangle$. The graph was obtained by 
saturating the ten star-diagrams corresponding to the ten trace terms,
namely with $n_1=3$ univalent vertices, $n_2=2$ bi-valent ones,
$n_3=1$ tri-valent one, $n_4=2$ four-valent ones, $n_6=1$ 
six-valent one and $n_8=1$ eight-valent one, hence a total of $V=10$ vertices. 
This graph corresponds to one
particular Wick pairing for which we have drawn the $E=16$ connecting edges,
giving rise to $F=2$ oriented loops bordering the faces of $\Gamma$.}{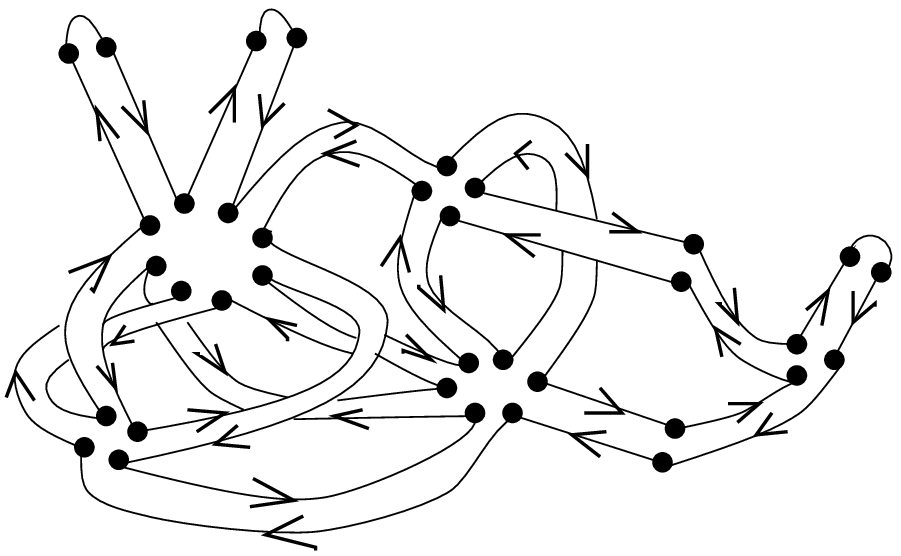}{8.cm}
\figlabel\gravi
\noindent In \gram,
we have first represented pictorially the integrand $\prod_i ({\rm
Tr}(M^i))^{n_i}$  as a succession of $n_i$ $i$-valent star diagrams like that
of \tracrep, $i=1,2,...$. Then we have summed 
over all possible saturations of all the marked half-edges of all these stars, 
thus forming (non-necessarily connected) ribbon or fatgraphs
$\Gamma$ with some labelling of their half-edges (see Fig.\gravi\ for an example 
of connected fatgraph). 
In \gram, we have  denoted by
$E(\Gamma)$ the total number of edges of $\Gamma$, connecting
half-edges by pairs, i.e. the number of propagators needed (yielding a factor
$1/N$ each, from \prop). The number $F(\Gamma)$ is the total number of faces of 
$\Gamma$. The faces of $\Gamma$ are indeed well-defined because $\Gamma$ is
a fatgraph, i.e. with edges made of doubly oriented parallel lines carrying the
corresponding matrix indices $i=1,2,...N$: the oriented loops we have
created by the pairing process are interpreted as face boundaries, in
one-to-one correspondence with faces of $\Gamma$. But the traces of the various
powers of $M$ still have to be taken, which means all the indices running from
$1$ to $N$ have to be summed over all these loops. This results in the factor
$N$ per face of $\Gamma$ in \gram. Finally, the  sum extends over all (possibly
disconnected) fatgraphs $\Gamma$ with labelled half-edges. Each such labelled
graph corresponds to exactly one Wick  pairing of \wimat. Summing over all the
possible labellings of a given un-labelled fatgraph $\Gamma$ results in some
partial cancellation of the symmetry prefactors $\prod_i 1/(i^{n_i}n_i!)$,
which actually leaves us with the inverse of the order of the symmetry group
of the un-labelled fatgraph $\Gamma$, denoted by $1/|Aut(\Gamma)|$. This gives
the final form
\eqn\fatag{ Z_N(g_1,g_2,...)=\sum_{{\rm fatgraphs}\atop \Gamma } \ 
{N^{V(\Gamma)-E(\Gamma)+F(\Gamma)}
\over |Aut(\Gamma)|}  
\prod_{i\geq 1} g_i^{n_i(\Gamma)} }
where $n_i(\Gamma)$ denotes the total number of $i$-valent vertices of $\Gamma$
and $V(\Gamma)=\sum_i n_i(\Gamma)$ is the total number of vertices of $\Gamma$.
To restrict the sum in \fatag\ to only connected graphs, we simply have to
formally expand the logarithm of $Z_N$, resulting in the final identity
\eqn\parfure{ F_N(g_1,g_2,...)={\rm Log}\, Z_N(g_1,g_2,...)=
\sum_{{\rm connected}\atop {\rm fatgraphs}\ \Gamma} 
{N^{2-2h(\Gamma)} \over |Aut(\Gamma)|}
\prod_i g_i^{n_i(\Gamma)} }
where we have identified the Euler characteristic $\chi(\Gamma)=F-E+V=2-2h(\Gamma)$,
where $h(\Gamma)$ is the genus of $\Gamma$ (number of handles).
Eqn.\parfure\ gives a clear geometrical meaning to the Gaussian average of our
choice of $f(M)$: it amounts to computing the generating function for fatgraphs
of given genus and given vertex valencies. 
Such a fatgraph $\Gamma$ is in turn dual to a 
tessellation $\Gamma^*$ of a Riemann surface of same genus,
by means of $n_i$ $i$-valent polygonal tiles, $i=1,2,...$. 

The result \parfure\ is therefore a statistical sum over discretized random
surfaces (the tessellations), that can be interpreted in physical terms as the
free energy of 2D quantum gravity. The name free energy stands generically for the
logarithm of the partition function $Z_N$. It simply identifies the Gaussian
matrix integral with integrand $f(M)$ as a discrete sum over configurations of
tessellated surfaces of arbitrary genera, weighted by some exponential 
factor. More precisely, imagine only $g_3=g\neq 0$ while all other $g_i$'s vanish.
Then \parfure\ becomes a sum over fatgraphs with cubic vertices (also called
$\phi^3$ fatgraphs), dual to triangulations $T$ of Riemann surfaces of arbitrary
genera. Assuming these triangles have all unit area, then $n_3(\Gamma)=A(T)$ is
simply the total area of the triangulation $T$. Hence \parfure\ becomes
\eqn\furpar{ F_N(g)=\sum_{{\rm connected}\ {\rm triangulations}\ T} 
{g^{A(T)} N^{2-2h(T)}\over |Aut(T)|} }
and the summand $g^A N^{2-2h}=e^{-S_E}$ is nothing but the exponential of the 
discrete version of Einstein's
action for General Relativity in 2 dimensions, 
which reads, for any surface $S$
\eqn\einstein{\eqalign{ S_E(\Lambda,{\cal N}\vert S)&= \Lambda \int_{S}
d^2x \sqrt{|g|} +{{\cal N} \over 4\pi} \int_{S} d^2x \sqrt{|g|} R \cr
&=\Lambda A(S) + {\cal N} (2-2h(S)) \cr}}
where $g$ is the metric of $S$, $R$ its scalar curvature, and
the two multiplicative constants are respectively
the cosmological constant $\Lambda$ and the Newton constant $\cal N$. 
In \einstein, we have identified the two invariants of $S$: its area $A(S)$ and its
Euler characteristic $\chi(S)=2-2h(S)$ (using the Gauss-Bonnet formula). 
The contact
with \furpar\ is made by setting $g=e^{-\Lambda}$ and $N=e^{-{\cal N}}$.
If we now include all $g_i$'s in \parfure\ we simply get a more elaborate
discretized model, in which we can keep track of the valencies of vertices of
$\Gamma$ (or tiles of the dual $\Gamma^*$).

Going back to the purely mathematical interpretation of \parfure, we start to
feel how simple matrix integrals can be used as tools for generating 
all sorts of graphs whose duals tessellate surfaces of arbitrary given 
topology. The size $N$ of the matrix relates to the genus, whereas the
details of the integrand relate to the structure of vertices. An important 
remark is also that the large $N$ limit of \parfure\ extracts the genus zero
contribution, namely that coming from planar graphs. So as a by-product,
it will be possible to extract from asymptotics of matrix integrals for large
size $N$ some results on planar graphs.

\subsec{Multi-Hermitian Matrix Case}

The results of previous section can be easily generalized to multiple Gaussian
integrals over several Hermitian matrices. More precisely, let $M_1$, $M_2$,
... $M_p$ denote $p$ Hermitian matrices of same size $N\times N$, and
$Q_{a,b}$, $a,b=1,2,...,p$ the elements of a positive definite form $Q$.
We consider the multiple Gaussian integrals of the form
\eqn\fromult{ \langle f(M_1,...,M_p)\rangle ={\int dM_1...dM_p e^{-{N\over 2}
\sum_{a,b=1}^p {\rm Tr}(M_a Q_{ab}M_b) } f(M_1,...,M_p) \over 
\int dM_1...dM_p e^{-{N\over 2}
\sum_{a,b=1}^p {\rm Tr}(M_a Q_{ab}M_b) } } }
where by a slight abuse of notation we still denote averages by the same
bracket sign as before: the one-Hermitian matrix case of the previous section
corresponds simply to $p=1$ and $Q_{1,1}=1$. The averages \fromult\
are computed by extending the
source integral method of previous section: for some Hermitian matrices
$S_1,...,S_p$ of size $N\times N$, we define and compute the
multi-source integral 
\eqn\soucmu{\Sigma(S_1,...,S_p)= 
\langle e^{\sum_{a=1}^p {\rm Tr}(S_a M_a)} \rangle =
e^{{1 \over 2N} \sum_{a,b=1}^p {\rm Tr}(S_a (Q^{-1})_{a,b} S_b)} }
and use multiple derivatives thereof to compute any expression of the form
\fromult, before taking $S_a\to 0$. As before, derivatives wrt elements of the
$S$'s must go by pairs to yield a non-zero result. For instance, in the case
of two matrix elements of $M$'s we find the propagators
\eqn\propmu{ \langle (M_a)_{ij} (M_b)_{kl}\rangle ={1\over N} \delta_{il} \delta_{jk} 
(Q^{-1})_{a,b} }
In general we will apply the multimatrix Wick theorem
\eqn\muwic{ \langle \prod_{(a,i,j)\in J} (M_a)_{ij} \rangle=
\sum_{{\rm pairings}\atop P} \prod_{{\rm pairs}\atop
(aij),(bkl)\in P} \langle (M_a)_{ij} (M_b)_{kl}\rangle }
expressing the multimatrix Gaussian average of any product of matrix elements of the
$M$'s as a sum over all pairings saturating the matrix half-edges, weighted 
by the corresponding value of the propagator \propmu. Note that half-edges 
must still be connected according to the rule \propag, but that in addition,
depending on the form of $Q$, some matrices may not be allowed to connect 
to one another (e.g. if $(Q^{-1})_{ab}=0$ for some $a$ and $b$,
then $\langle M_a M_b\rangle=0$,
and there cannot be any edge connecting a matrix with index $a$ to one
with index $b$).

This gives us much freedom in ``cooking up" multimatrix models to evaluate generating 
functions of graphs with specific decorations such as colorings, spin models,
etc... This is expected to describe the coupling of matter systems (e.g.
a spin model usually defined on a regular lattice) to 2D quantum gravity (by
letting the lattice fluctuate into tessellations of arbitrary genera).

An important example is the so-called gravitational $O(n)$ model \ON, say on cubic
fatgraphs. Its regular lattice version \NIN\ is as spin model 
defined on the honeycomb (hexagonal) 
lattice, with only cubic (trivalent) vertices. A spin configuration of the model is simply a map
from the set of vertices of the lattice to the $n$-dimensional unit sphere, hence
the name inherited from the obvious symmetry of the target space. 
For $n=1$, this is the Ising model, in which the spin variable may only take
values $\pm 1$.
For $n\geq 2$ this
gives an infinite continuum set of configurations, over which we have to perform
a statistical sum, henceforth an integration over all the spin values. With a suitable
choice of nearest neighbor interaction between adjacent spins, this statistical sum
was shown to reduce to the following very simple loop model on the honeycomb lattice.
The partition function $Z_{O(n)}$ of the regular lattice model is expressed as 
a sum over configurations of closed non-intersecting loops drawn on the honeycomb lattice, 
each weighed by the factor $n$, whereas each edge of each loop receives a 
Boltzmann factor $K$ corresponding to the inverse temperature of the statistical
mechanical model. The gravitational version thereof is very simple to figure out:
we must replace the honeycomb lattice by its spatial fluctuations into arbitrary
cubic fatgraphs. Then, on each fatgraph, we must consider arbitrary closed
non-intersecting loop configurations, with the same weights as in the regular lattice
case. The free energy of this model reads
\eqn\freon{ F_N(g\vert K,n)=  \sum_{{\rm connected}\atop {\rm cubic}\ {\rm
fatgraphs}\ \Gamma} {g^{A(\Gamma)}N^{2-2h(\Gamma)} \over
|Aut(\Gamma)|}\sum_{{\rm loop}\atop {\rm configurations}\ \ell}
n^{C(\ell)} K^{E(\ell)} }
where as before we  have inserted a weight $g$ per cubic vertex ($A(\Gamma)$
is the total area of the triangulation dual to $\Gamma$, also the total number
of vertices of $\Gamma$), and for each loop configuration $\ell$ on $\Gamma$,
we have denoted by $C(\ell)$ and $E(\ell)$ respectively the total number of connected
components of $\ell$ (number of loops) and the total number of edges occupied by
the loops.
Let us now present a multi-matrix model for \freon. Loop models can be easily
represented by using the so-called {\it replica trick}: as we wish to attach
a weight $n$ per loop, we may introduce a color variable $c=1,2,...,n$ by which
we may paint each connected component of loop independently. Summing over all
possible colorings will yield the desired weight. Hence we are allowing the loops
to be {\it replicated}, namely we may now choose among $n$ different types of loops 
to build our configurations. To represent these configurations, we need 
$n+1$ Hermitian matrices of same size $N\times N$, say $A_1,A_2,...,A_n$ for
the $n$ colored loop half-edges, and $B$ for the unoccupied half-edges. 
As loops are non-intersecting the only allowed vertices are
of the form $B^3$ (unoccupied vertex) and $A_c^2 B$ 
(loop of color $c$ going through a vertex, with exactly one unoccupied edge).
The propagators must force loops to conserve the same color on each connected
component, while unoccupied half-edges can only be connected to unoccupied ones,
namely
\eqn\propaga{\eqalign{ \langle (A_c)_{ij} (A_d)_{kl}\rangle 
&={K\over N}\delta_{cd}\delta_{il}\delta_{jk}\cr
\langle B_{ij} B_{kl}\rangle &= {1\over N}\delta_{il}\delta_{jk}\cr
\langle  (A_c)_{ij}  B_{kl}\rangle &= 0\cr}}
where we also have added the weight $K$ per edge of loop.
This defines the inverse of the quadratic form $Q$ we need for the corresponding
matrix model, according to \propmu. This leads to the multi-matrix integral
\eqn\onmat{\eqalign{
Z_N(g\vert K,n)&={ \int dB dA_1...dA_n e^{-N{\rm Tr}(V(B,A_1,...,A_n))}
\over \int  dB dA_1...dA_n e^{-N{\rm Tr}(V_0(B,A_1,...,A_n))} }\cr
V(B,A_1,...,A_n)&={1\over 2}B^2+ {1\over 2K}\sum_{c=1}^n A_c^2
-{g\over 3} B^3-gB(\sum_{c=1}^n A_c^2)\cr
V_0(B,A_1,...,A_n)&={1\over 2}B^2+ {1\over 2K}\sum_{c=1}^n A_c^2 \cr}}
To get the free energy of the gravitational $O(n)$ model \freon, we simply have
to take the logarithm of \onmat, to extract the contribution
from connected fatgraphs only. As before, we may decorate the model with 
higher order terms of the form $\sum_i g_i B^i/i$ to allow for
arbitrary unoccupied vertex configurations and keep track of the fine
structure of the fatgraphs, without altering  the loop configurations. Another
possibility is to remove the $B^3$ term from \onmat, to restrict the loop
configurations to compact ones, i.e. covering all the vertices of $\Gamma$.

Another standard example, with a more complicated quadratic form, 
is the so-called gravitational $q$-states Potts
model \QPO, defined say on arbitrary cubic diagrams $\Gamma$. Imagine that the
diagram $\Gamma$ is decorated by maps $\sigma$ from its set of vertices
$v(\Gamma)$ to $\IZ_q$. We may alternatively view this as ``spin" (or color)
variables $\sigma\in \IZ_q$ living on the vertices of $\Gamma$. The model
is then defined by attaching
Boltzmann factors corresponding to different energies of configurations of
neighboring spins according to whether they are identical or distinct. More
precisely, if $v,v'$ are two vertices connected by an edge $e$ in $\Gamma$, we
define the weight 
\eqn\bowpot{ w_e(\sigma)=e^{K \delta_{\sigma(v),\sigma(v')}}} 
for some positive real parameter $K$ (inverse temperature).
The gravitational Potts model free energy is then defined as
\eqn\frepot{ F_N(g\vert K,q)= \sum_{{\rm connected}\atop {\rm cubic}\ {\rm
fatgraphs}\ \Gamma} {g^{A(\Gamma)}N^{2-2h(\Gamma)} \over
|Aut(\Gamma)|}\sum_{{\rm maps}\atop \sigma:v(\Gamma)\to \IZ_q} \prod_{{\rm edges}\atop
e\ {\rm of}\ \Gamma} w_e(\sigma) }
Let us now ``cook-up" a multi-matrix model integral for \frepot.
We wish to generate all connected cubic fatgraphs $\Gamma$: this 
would be easily done by considering the example of the previous section
with all $g$'s equal to zero except $g_3=g$. To represent the spin
configurations however, we need to be able to distinguish between
$q$ different types
of vertices, according to the value of $\sigma(v)\in \IZ_q$. This is done by
introducing $q$ Hermitian matrices $M_1,...,M_q$ with the same size $N\times
N$, with a Gaussian potential of the form $N$Tr$(\sum M_a Q_{ab}M_b)/2$,
such that
\eqn\condiq{ (Q^{-1})_{ab} = 1+ (e^K-1) \delta_{ab} }
in order for the propagators \propmu\ to receive the extra weight $e^K$ when
the ``spins" $a$ and $b$ are identical, and $1$ when they are distinct.
Eqn.\condiq\ is easily inverted into
\eqn\valq{ Q_{ab}= {1\over e^K-1} ( \delta_{ab} -{1\over e^K-1+q} ) }
The free energy of the gravitational Potts model on cubic diagrams
is therefore given by
\eqn\grpo{ \eqalign{
F_N(g\vert K,q)&= {\rm Log}\, Z_N(g\vert K,q) \cr
Z_N(g \vert K,q)&={\int dM_1...dM_q e^{-{N}{\rm Tr}V(M_1,...,M_q)} \over
\int dM_1...dM_q e^{-{N}{\rm Tr}V_0(M_1,...,M_q)} }\cr
V(M_1,...,M_q)&= {1\over 2} \sum_{a,b=1}^q Q_{ab} M_aM_b -{g\over 3}
\sum_{a=1}^q M_a^3 \cr
V_0(M_1,...,M_q)&= {1\over 2} \sum_{a,b=1}^q Q_{ab} M_aM_b \cr}}
with $Q$ as in \valq. When $q=2$ the model is nothing but the Ising model
(also equivalent to the $O(n=1)$ model, as mentioned above).
To recover \frepot\ from \grpo, we simply apply the multi-matrix Wick theorem,
and note that the overall contribution for each graph is
$N^{V-E+F}=N^{2-2h}$ as before, as we get a weight $N$ per vertex, $1/N$ per
edge from propagators, and $N$ per oriented loop of indices, irrespectively
of the spin values.

\subsec{A generating Function for Fatgraphs}

In this section, we use the philosophy developed in the two previous ones 
to present a model of {\it dually weighted} fatgraphs, namely in which 
we wish to specify not only the structure of vertices but that of faces
as well \FAT. This uses the slightly more general notion of index-dependent
propagators, obtained by considering one-Hermitian matrix models with 
Gaussian potentials of the form
\eqn\gaupol{ V(M)= {1\over 2} \sum_{ijkl} M_{ij} Q_{ij;kl} M_{kl} }
where $Q$ is some arbitrary tensor depending on the four 
matrix indices, but such that Tr$V(M)$ remains positive. Repeating the
calculations of Sect. 2.3, we find propagators of the form
\eqn\propnew{ \langle M_{ij} M_{kl} \rangle = {1\over 2N} ({1\over Q_{ij;kl}}+
{1\over Q_{kl;ij}}) }
This gives even more freedom in tailoring specific models for combinatorial
purposes. Let us now examine the choice
\eqn\choiq{ Q_{ij;kl} = \lambda_i \lambda_j \delta_{il}\delta_{jk} }
for some $\lambda_1,...,\lambda_N>0$. We will also refer to the $N\times N$
diagonal matrix
$\Lambda$ with entries $\Lambda_{ij}=\lambda_i \delta_{ij}$.
Then the propagator \propnew\ becomes
\eqn\becpro{ \langle M_{ij} M_{kl} \rangle = {1\over N\lambda_i\lambda_j}
\delta_{il}\delta_{jk} }
If we consider the Gaussian average wrt \gaupol\ of the function
$f(M)=\exp(-N\sum_{i\geq 1} g_i {\rm Tr}(M^i)/i)$, we find the following
interpretation after taking the logarithm:
\eqn\loginter{\eqalign{F_N(\Lambda;g_1,g_2,...)&=
{\rm Log}\, \langle e^{-N\sum_{i\geq 1} {g_i\over i} {\rm Tr}(M^i)} \rangle
\cr
&= \sum_{{\rm connected}\atop {\rm fatgraphs}\ \Gamma}{N^{2-2h(\Gamma)} \over
|Aut(\Gamma)|} \prod_{i\geq 1} g_i^{n_i(\Gamma)}  \prod_{{\rm faces} 
\atop f\ {\rm of}\  \Gamma} \sum_{i=1}^N {1\over \lambda_i^{m(f)}} \cr}}
where, for each face $f$ of $\Gamma$, we have denoted by $m(f)$ the perimeter
of its boundary (number of adjacent edges, or valency). The latter term comes
from
the summation over all the indices running over the oriented loops of the
graph. But this time, we do not use $\sum_{i=1}^N 1=N$, but rather notice that 
each edge of a loop carrying the running index $i$ comes with a factor 
$1/\lambda_i$ from the propagators
\becpro. Hence each face comes with a total factor $1/\lambda_i^{m(f)}$,
$m(f)$ the face valency,  which
has to be summed over from $i=1$ to $N$, to yield \loginter.
Introducing the variables 
\eqn\miwak{ G_k= {\rm Tr} (\Lambda^{-k})=\sum_{i=1}^N {1 \over \lambda_i^k}}
we finally get the expansion
\eqn\expfn{ F_N(\Lambda;g_1,g_2,..)=\sum_{{\rm conn.}\
{\rm fatgraphs}\ \Gamma} {N^{2-2h(\Gamma)}\over |Aut(\Gamma)|}
\prod_{i\geq 1} g_i^{n_i(\Gamma)} G_i^{m_i(\Gamma)} }
where $m_i(\Gamma)$ denotes the number of $i$-valent faces of $\Gamma$.
The expression \expfn\ looks almost symmetric in $g_i$ and $G_i$: it
would actually be the case if the $G_i$ were an infinite set of independent 
variables like the $g_i$'s, but they are not. Due to the 
definition \miwak, only the $N$ first $G$'s are generically independent variables,
all the others are dependent. Here is how we should read \expfn. For
any given $N$, let us expand $F_N(\Lambda;g_1,g_2,...)$ as a sum over fatgraphs
with $\leq N$ edges. Then the rhs only involves the numbers $G_1,...,G_N$,
as well as $g_1,...,g_N$, and is manifestly symmetric in the $g$'s and $G$'s,
as the two sets of variables are exchanged by going to the dual graphs. 
The coefficient of each monomial corresponds to the
graphs with specified genus, numbers of $i$-valent vertices, and numbers
of $i$-valent faces. To get information on larger graphs, we simply have to
take $N$ larger and larger.

\newsec{Matrix Models: Solutions}

We hope that this series of examples has convinced the reader of the relevance
of matrix integrals to graph combinatorics. Let us now see how to extract
useful information from these matrix integrals. In this section, we will
mainly cover the one-matrix integrals defined in Sect. 2.2. Multi-matrix
techniques are very similar, and we will present them later, when we need them.
More precisely, we will study the one-matrix integral
\eqn\onmav{ Z_N(V)={\int dM e^{-N{\rm Tr}\, V(M)} \over
\int dM e^{-N{\rm Tr}\, V_0(M)} }}
with an arbitrary polynomial potential 
$V(x)={x^2\over 2}+ \sum_{3\leq i\leq k} {g_i\over i} x^i$ and $V_0(x)={x^2\over 2}$.
This contains as a limiting case the partition function \gram\ of Sect. 2.2.
Note also that we are not worrying at this point about convergence issues\foot{
The reader may be more comfortable assuming Re$(g_k)>0$ to actually ensure
convergence throughout this section.} for these
integrals, as they must be understood as formal tools allowing for computing
well-defined coefficients in formal series expansions in the $g$'s. 

\subsec{Reduction to Eigenvalues}

The step zero in computing the integral \onmav\ is the reduction to $N$ one-dimensional 
integrals, namely over the real eigenvalues $m_1,...,m_N$ of the Hermitian matrix $M$.
This is done by performing the change of variables $M\to (m,U)$, where 
$m=$diag$(m_1,...,m_N)$, and $U$ is a unitary diagonalization matrix such that
$M=UmU^\dagger$, hence $U\in U(N)/U(1)^N$ as $U$ may be multiplied by an arbitrary
matrix of phases. The Jacobian of the transformation is readily found to be
the squared Vandermonde determinant
\eqn\jaco{ J= \Delta(m)^2 = \prod_{1\leq i<j\leq N} (m_i-m_j)^2 }
Performing the change of variables in both the numerator and denominator of \onmav\ 
we obtain
\eqn\obtz{ Z={\int_{\IR^N} dm_1...dm_N \Delta(m)^2 e^{-N\sum_{i=1}^N V(m_i)}
\over \int_{\IR^N} dm_1...dm_N \Delta(m)^2 e^{-N\sum_{i=1}^N {m_i^2\over 2}} } }

\subsec{Orthogonal Polynomials}

The standard technique of computation of \obtz\ uses orthogonal polynomials.
The idea is to disentangle the Vandermonde determinant squared interaction
between the eigenvalues.
The solution is based on the following simple lemma: 
if $p_m(x)=x^m+\sum_{j=0}^{m-1} p_{m,j}x^j$
are monic polynomials of degree $m$, for $m=0,1,...,N-1$, then
\eqn\vandereq{ \Delta(m)=\det(m_i^{j-1})_{1\leq i,j \leq N} =
\det(p_{j-1}(m_i))_{1\leq i,j \leq N} }
easily derived by performing suitable linear combinations of columns.
Let us now introduce the unique set of monic polynomials $p_m$, 
of degree $m=0,1,...,N-1$, that are
orthogonal wrt the one-dimensional
measure $d\mu(x)=\exp(-NV(x))dx$, namely such that
\eqn\monort{(p_m,p_n)= \int_{\IR} p_m(x)p_n(x) d\mu(x) = h_m \delta_{m,n} }
These allow us to rewrite the numerator of \obtz, using \vandereq,  as
\eqn\numera{ \sum_{\sigma,\tau \in S_N}
\epsilon(\sigma \tau) \prod_{i=1}^N \int_{\IR} d\mu(m_i) p_{\sigma(i)-1}(m_i)
p_{\tau(i)-1}(m_i) e^{-NV(m_i)} = N! \prod_{j=0}^{N-1} h_j }
We may apply the same recipee to compute the denominator, with the result
$N! \prod_{j=0}^{N-1} h_j^{(0)}$, where the $h_j^{(0)}$ are the squared norms
of the orthogonal polynomials wrt to the Gaussian measure $d\mu_0(x)=\exp(-Nx^2/2)dx$.
Hence the $h$'s determine $Z_N(V)$ entirely through
\eqn\thru{ Z_N(V)=\prod_{i=0}^{N-1} {h_i\over h_i^{(0)} } }

We may repeat this calculation in the presence of a ``spectator" term
of the form $\sum_i f(m_i)/N$ for some arbitrary function $f$. The term corresponding
to $i=1$ in this sum simply reads, by obvious modification of \numera\
\eqn\repla{\eqalign{  \sum_{\sigma,\tau \in S_N}
\epsilon(\sigma \tau) \int_{\IR} &d\mu(m_1) {f(m_1)\over N} p_{\sigma(1)-1}(m_1)
p_{\tau(1)-1}(m_1) e^{-NV(m_1)} \times \cr
&\times \prod_{i=2}^N \int_{\IR} d\mu(m_i) p_{\sigma(i)-1}(m_i)
p_{\tau(i)-1}(m_i) e^{-NV(m_i)}\cr
&= N! \prod_{j=0}^{N-1} h_j\times {1\over N}\int_\IR dm f(m) \sum_{i=0}^{N-1} {p_i(m)^2\over N h_i}
e^{-NV(m)} \cr} }
which is independent of the choice of index $i$ ($=1$ here). Therefore, the complete average
wrt to the full matrix measure reads
\eqn\fulav{\eqalign{ 
\langle {1\over N}\sum_{i=1}^N f(m_i)\rangle_V &= {\int_{\IR^N} dm_1...dm_N  
{1\over N}\sum_{i=1}^N f(m_i) \Delta(m)^2 e^{-N\sum V(m_i)} \over 
\int_{\IR^N} dm_1...dm_N  \Delta(m)^2 e^{-N\sum V(m_i)}} \cr
&=  \int_\IR dm f(m) \sum_{i=0}^{N-1} {p_i(m)^2\over N h_i}
e^{-NV(m)} \cr} }
where we have added the subscript $V$ to recall that the average is normalized wrt
the full measure (using $V$ instead of $V_0$ in the denominator term).
In the case where $f(m)$ is a polynomial, this can be immediately rewritten in terms of the
$r$'s and therefore the $h$'s only, as we'll explain now.

To further compute the $h$'s, let us introduce the two following operators $Q$ and
$P$, acting on the polynomials $p_m$:
\eqn\opq{ \eqalign{ Qp_m(x)&= x p_m(x) \cr
Pp_m(x) &= {d \over dx} p_m(x) \cr}}
with the obvious commutation relation 
\eqn\canocom{ [P,Q]=1}
Using the self-adjointness of $Q$ wrt the scalar product $(f,g)=\int f(x)g(x) d\mu(x)$,
it is easy to prove that 
\eqn\qrec{ Qp_m(x)=xp_m(x)=p_{m+1}(x)+s_m p_m(x)+r_m p_{m-1}(x) }
for some constants $r_m$ and $s_m$,
and that $s_m=0$ if the potential $V(x)$ is even. 
The same reasoning yields
\eqn\rsm{ r_m={h_m \over h_{m-1}}, \ \ m=1,2,...} 
and we also set $r_0=h_0$ for convenience.
 
Moreover, expressing both $(Pp_m,p_m)$ and $(Pp_m,p_{m-1})$ in two ways, 
using integration by parts, we easily get the master equations
\eqn\masteq{\eqalign{ {m\over N} &= {(V'(Q)p_m,p_{m-1})\over (p_{m-1},p_{m-1})} \cr
0&=(V'(Q)p_m,p_m) \cr}}
which amount to a recursive system for $s_m$ and $r_m$. Note that the second line 
of \masteq\ is automatically satisfied if $V$ is even: it vanishes as the integral
over $\IR$ of an odd function. Indeed, this latter equation allows for computing the $s$'s
out of the $r$'s, therefore becomes a tautology when all the $s$'s vanish.
Assuming for simplicity that $V$ is even, the first equation of \masteq\ gives
a non-linear recursion relation for the $r$'s. The degree $k$ of $V$ actually determines the 
number of terms in the recursion, namely $k-1$. So, we need to feed the $k-2$ initial values 
of $r_0,r_1,r_2,...,r_{k-3}$ into the recursion relation, and we obtain the exact value of
$Z_N(V)$ by substituting $h_i=r_0r_1...r_i$ in both the numerator and the denominator
of \thru. Note that for $V_0(x)=x^2/2$ the recursion \masteq\ reduces simply to
\eqn\zercas{ {m\over N}= {(Qp_m^{(0)},p_{m-1}^{(0)})\over (p_{m-1}^{(0)},p_{m-1}^{(0)})}
= r_m^{(0)} }
and therefore $h_m^{(0)}=h_0^{(0)} m!/N^m =\sqrt{2\pi} m!/N^{m+1/2}$. The $p_m^{(0)}$ are 
simply the Hermite polynomials.  

Finally, the free energy of the model \onmav\ reads
\eqn\frenen{ F_N(V)={\rm Log}\, Z_N(V)=N\,{\rm Log}\,r_0\sqrt{N\over 2\pi}\, + 
\sum_{i=1}^{N-1} (N-i){\rm Log} {Nr_i\over i}} 
in terms of the $r$'s. In the case when $V$ is even and $f(m)$ a polynomial, 
the average \fulav\ is clearly expressible in terms of the $r$'s only, by use of 
the recursion relation \qrec\ (with $s_m=0$). Indeed, when $f(m)=m^n$, we get
\eqn\recuvelo{ \langle {1\over N}{\rm Tr}(M^n) \rangle_V ={1\over N}\sum_{i=0}^{N-1}
{(Q^np_i,p_i)\over h_i}} 
When $n=2$ for instance, this simply reads
\eqn\sitwo{  \langle {1\over N}{\rm Tr}(M^2) \rangle_V ={r_1\over N}+\sum_{i=1}^{N-1}
{r_{i+1}+ r_i\over N} }

\subsec{Large $N$ asymptotics I: Orthogonal Polynomials}

As mentioned before, the large $N$ limit of matrix integrals always has an interpretation
as sum over planar (genus zero) fatgraphs. In particular, the free energy $F_N(V)$ of
\frenen\ can be expressed in an analogous way  as \parfure\ as a sum over connected
fatgraphs, except that all but a finite number of $g$'s vanish, namely $g_m=0$ for
$m\geq k+1$, and also for $m\leq 2$. The large $N$ contribution is 
\eqn\larn{ F_N(V)\sim N^2 f_0(V) +O(1) }
where $f_0(V)$ is the planar free energy, namely that obtained by restricting oneself
to genus zero fatgraphs.

In view of the expression \frenen, it is straightforward to get asymptotics like 
\larn, by first noting that as $h_0\sim \sqrt{2\pi \over N}$,
the first term in \frenen\ doesn't contribute to the leading order $N^2$
and then by approximating the sum by an integral of the form
\eqn\intaprox{ f_0(V)=\lim_{N\to \infty} {1\over N} \sum_{i=1}^{N-1}
(1-{i\over N}) {\rm Log} {r_i\over i/N} = \int_0^1 dz (1-z) {\rm Log}{r(z)\over z}}
where we have assumed that the sequence $r_i$ tends to a function $r_i\equiv r(i/N)$ of
the variable $z=i/N$ when $N$ becomes large. This can actually be proved rigorously.
The limiting function $r(z)$ in \intaprox\ is determined by the equations
\masteq, that become polynomial in this limit. In the case $V$ even for 
instance, where $V(x)=x^2/2+ \sum_{i=2}^p g_{2i}x^{2i}/(2i)$ ($k=2p$), we simply get
\eqn\req{ z= r(z)+\sum_{i=2}^p {2i-1 \choose i} g_{2i} r(z)^i }
The function $r(z)$ is the unique root of this polynomial equation that tends to $z$
for small $z$ (it can be expressed using the Lagrange inversion method for instance,
as a formal power series of the $g$'s), and the free energy follows from \intaprox.  
This allows also for computing the large $N$ limit of averages of the form \fulav\ or
\recuvelo. For instance, the large $N$ limit of \sitwo\ reads
\eqn\sitoN{ \lim_{N\to \infty} \langle  {1\over N}{\rm Tr}(M^2) \rangle_V =
2\int_0^1 dz\, r(z) }
and more generally we have
\eqn\sitotN{ \lim_{N\to \infty} \langle  {1\over N}{\rm Tr}(M^n) \rangle_V =\left\{
\matrix{ {2p\choose p}\int_0^1 dz\, r(z)^p & {\rm if} \ n=2p \cr
0 & {\rm otherwise} \cr} \right. }
Note that in the case $V=V_0$ (and $r(z)=z$) of Gaussian averages, \sitotN\
reduces to the result \magau. 

\subsec{large $N$ asymptotics II: Saddle-Point Approximation}

Let us now present another solution, which does not rely on the orthogonal
polynomial technique nor requires its applicability. We start from the 
$N$-dimensional integrals \obtz, that we rewrite
\eqn\robtz{ Z_N(V)={\int dm_1...dm_N e^{-N^2 S(m_1,...,m_N)} \over
\int dm_1...dm_N e^{-N^2 S_0(m_1,...,m_N)}} }
where we have introduced the ``actions"
\eqn\actio{ \eqalign{
S(m_1,...,m_N)&= {1 \over N}\sum_{i=1}^NV(m_i) -{1\over N^2}\sum_{1\leq i\neq j\leq N}
{\rm Log}\vert m_i-m_j\vert \cr
S_0(m_1,...,m_N)&={1 \over N}\sum_{i=1}^NV_0(m_i) -{1\over N^2}\sum_{1\leq i\neq j\leq N}
{\rm Log}\vert m_i-m_j\vert \cr}}
For large $N$ the numerator and denominator of \robtz\ are dominated by the 
semi-classical (or saddle-point) minimum of $S$ and $S_0$ respectively.
For $S$, the saddle-point equations read
\eqn\sap{ {\partial S\over\partial m_j}=0 \ \Rightarrow \ 
V'(m_j)= {2 \over N} \sum_{1\leq i \leq N\atop i\neq j} {1 \over m_j-m_i} }
for $j=1,2,...,N$.
Introducing the discrete resolvent 
\eqn\dires{ \omega_N(z)={1\over N} \sum_{i=1}^N {1\over z-m_i} }
and multiplying \sap\ by $1/(N(z-m_j))$ and summing over $j$, 
we easily get the equation
\eqn\qares{\eqalign{
V'(z) \omega_N(z)&+{1\over N}\sum_{j=1}^N {V'(m_j)-V'(z)\over z-m_j}\cr
&={1\over N^2} \sum_{1\leq i\neq j\leq N} 
{1\over m_j-m_i}\bigg({1\over z-m_j}-{1\over z-m_i}
\bigg) \cr
&={1\over N^2} \sum_{1\leq i\neq j\leq N} {1\over (z-m_i)(z-m_j)}\cr
&=\omega_N(z)^2+{1\over N}\omega_N'(z) \cr}}
Assuming $\omega_N$ tends to a differentiable function $\omega(z)$ when $N\to \infty$ 
we may neglect the last derivative term, and we are left with the quadratic equation
\eqn\qadrares{\eqalign{ &\omega(z)^2 -V'(z) \omega(z)+ P(z) =0 \cr
&P(z)=\lim_{N\to \infty}{1\over N}\sum_{j=1}^N {V'(z)-V'(m_j)\over z-m_j} \cr}}
where $P(z)$ is a polynomial of degree $k-2$. The existence of the limiting resolvent
$\omega(z)$ boils down to that of the limiting density of distribution of
eigenvalues
\eqn\lidens{ \rho(z)=\lim_{N\to \infty} {1\over N}\sum_{j=1}^N \delta(z-m_j)}
normalized by the condition 
\eqn\normaro{ \int_\IR \rho(z) dz =1 }
as there are exactly $N$ eigenvalues on the real axis. This density is related 
to the resolvent through
\eqn\resden{ \omega(z)=\int {\rho(x)\over z-x} dx=
\sum_{m=1}^\infty {1\over z^m} \int_\IR x^{m-1}\rho(x) dx } 
where the expansion holds in the large $z$ limit, and the integral extends over 
the support of $\rho$, included in the real line. 
Conversely, the density is obtained from the resolvent by use of the 
discontinuity equation across its real support 
\eqn\discon{ \rho(z)={1\over 2 i \pi}\lim_{\epsilon\to 0} \omega(z+i\epsilon)-
\omega(z-i\epsilon) \qquad z\in {\rm supp}(\rho) }
Solving the quadratic equation \qadrares\ as
\eqn\solqa{ \omega(z)= {V'(z)-\sqrt{ (V'(z))^2 -4 P(z)} \over 2} }
we must impose the large $z$ behavior inherited from \normaro\-\resden, namely
that $\omega(z)\sim 1/z$ for large $z$. For $k\geq 2$, the polynomial in the square 
root has degree $2(k-1)$: expanding the square root for large $z$ up to order $1/z$,
all the terms cancel up to order $0$ with $V'(z)$,
and moreover the coefficient in front of $1/z$ must be $1$ (this fixes the leading 
coefficient of $P$). The other coefficients of $P$ are fixed by the higher moments of
the measure $\rho(x) dx$. 
For instance, when $k=2$ and $V=V_0$, we get
\eqn\rezer{ \omega_0(z)= {1\over 2} (z-\sqrt{z^2-4} ) }
It then follows from \discon\ that the density has the compact support $[-2,2]$ and has the
celebrated ``Wigner's semi-circle law" form
\eqn\forro{ \rho_0(z)= {1\over 2\pi } \sqrt{4-z^2} } 
Viewing the resolvent $\omega_0$ as the generating function for the moments of the 
measure whose density is $\rho_0$ (through the expansion \resden), we immediately get
the values of the moments 
\eqn\mom{ \int_\IR x^n \rho_0(x) dx = \left\{ \matrix{ c_p & {\rm if}\ n=2p \cr
0 & {\rm otherwise} \cr} \right. }
with $c_p$ as in \catalan. These are indeed immediately identified
with the planar limit of the Gaussian Hermitian matrix averages (with potential $V_0(x)=x^2/2$)
by using the following identity, valid for any $V$
\eqn\idwi{ \lim_{N\to \infty}\langle {1\over N}{\rm Tr}M^n\rangle_{V}= \int_\IR 
x^n \rho(x) dx }
The result \mom\ agrees with \magau\ and \sitotN. 
Actually, comparing \idwi\ with \recuvelo, we deduce the following expression for the 
limiting density $\rho(x)$ in terms of the orthogonal polynomials of the previous
solution
\eqn\limidpo{ \rho(x)=\lim_{N\to \infty} {1 \over N} \sum_{i=0}^{N-1} 
{p_i(x)^2\over h_i} e^{-NV(x)}}
When $V=V_0$, we recover from \limidpo\  the Wigner's semi-circle from standard asymptotics of the 
Hermite polynomials. 

In the general case, the density reads 
\eqn\geneden{ \rho(z)= {1\over 2\pi}\sqrt{4P(z)-(V'(z))^2} }
and may have a disconnected support, made of a union of intervals. 
It is however interesting to restrict oneself to the case when the support of $\rho$ is
made of a single real interval $[a,b]$. It means that the polynomial $V'(z)^2-4P(z)$ has 
single roots at $z=a$ and $z=b$ and that all other roots have even multiplicities.
In other words, we may write
\eqn\omez{\omega(z)={1\over 2}(V'(z) -Q(z)\sqrt{(z-a)(z-b)} )}
where $Q(z)$ is a polynomial of degree $k-2$, entirely fixed in terms of $V$ by the 
asymptotics $\omega(z)\sim 1/z$ for large $|z|$. 
For instance, for an even quartic potential, we have
\eqn\evenquart{\eqalign{
V(z)&={z^2\over 2}-g {z^4\over 4} \cr
\omega(z)&={1\over 2}(z-g z^3 -(1-g{a^2\over 2}-g z^2)\sqrt{z^2-a^2})\cr
a^2&={2\over 3g}(1-\sqrt{1-12 g})\cr
\rho(z)&={1\over 2 \pi}(1-g{a^2\over 2}-g z^2)\sqrt{a^2-z^2} \cr}}

The planar free energy \larn\
is finally obtained by substituting the limiting densities $\rho,\rho_0$ in the 
saddle point actions, namely
\eqn\frenden{\eqalign{ f_0(V)&=S(\rho_0,V_0)-S(\rho,V)\cr 
S(\rho,V)&=\int dx \rho(x) V(x) -\int dx dy \rho(x)\rho(y) {\rm Log}\vert x-y\vert\cr}}
This expression seems more involved than our previous result \intaprox, but is equivalent 
to it. In the case of the quartic potential of \evenquart, we find the genus zero
free energy $f_0(g)\equiv f_0(V)$
\eqn\genefg{ f_0(g)= {1\over 2}{\rm Log}\, {a^2\over 4} +{1\over 384}(a^2-4)(a^2-36) }
with $a^2$ as in \evenquart. 

In cases where the orthogonal polynomial technique does not apply
(like in complicated multi-matrix integrals), however, 
the saddle-point technique always gives acces to the planar limit.

For completeness let us quickly mention the corresponding results for the 
gravitational O(n) model introduced in Sect. 2.3 (see \ONSOL\ for details and a more
general solution).
We have to evaluate the large $N$ asymptotics of the partition function
\eqn\parfon{\eqalign{
Z&={\int dA_1...dA_n dB e^{-N{\rm Tr}(W(B,A_1,A_2,...,A_n))}\over
\int dA_1...dA_n dB e^{-N{\rm Tr}(W_0(B,A_1,A_2,...,A_n))}}\cr
W(B,A_1,A_2,...,A_n)&= V(B) +{1\over 2}\sum_{c=1}^n A_c^2 -gB\sum_{c=1}^n A_c^2 \cr
W_0(B,A_1,A_2,...,A_n)&={1\over 2} B^2 +{1\over 2}\sum_{c=1}^n A_c^2 \cr}}
a simple generalization of \onmat\ with some arbitrary potential $V(B)$.
Note that $W$ is simply quadratic in the $A_c$, so we can perform the Gaussian
integrals overs the $A$'s first. More precisely, the potential takes the
form
\eqn\potw{\eqalign{ {\rm Tr}\, W(B,A_1,A_2,...,A_n)&={\rm Tr}\, V(B)+{1\over 2}
\sum_{c=1}^n \sum_{ijkl=1}^N (A_c)_{ik} {\bf Q}_{ik;jl} (A_c)_{lj} \cr
{\bf Q}_{ik;jl} &= \delta_{ij}\delta_{kl} -g (\delta_{ij} B_{kl} +\delta_{kl} B_{ij}\cr}}
More compactly, the quadratic form reads
\eqn\quadon{ {\bf Q} = I\otimes I -g (I\otimes B+B \otimes I)}
The Gaussian integral over the $A$'s, normalized like in \parfon\ simply gives
det$({\bf Q})^{-1/2}$ for each integral, hence  
\eqn\zvalue{ Z={1\over \int dB e^{-N{\rm Tr}\,{B^2\over 2}} }\int dB e^{-N{\rm Tr}\, V(B)} 
\det(I\otimes I -g (I\otimes B+B \otimes I))^{-n/2} }
So the O(n) model reduces to a one-matrix model, but with a complicated potential.
We may now apply both the reduction to an eigenvalue integral and the
large $N$ technique sketched above. We get the eigenvalue integral
\eqn\eigint{\eqalign{
Z&={1\over (2\pi)^{N/2}}\int db_1 ... db_N e^{-N^2 S(b_1,...,b_N)} \cr
S(b_1,...,b_N)&= {1\over N}\sum_{i=1}^N V(b_i) 
-{1\over N^2}\sum_{1\leq i\neq j\leq N} {\rm Log}\,|b_i-b_j| \cr
&+{n\over 2N^2}
\sum_{1\leq i,j \leq N} {\rm Log} (1-g(b_i+b_j)) \cr}}
Note that in the last term we may impose the constraint $i\neq j$, as the 
terms $i=j$ only contribute for $O({1\over N})$ to $S$. The corresponding 
saddle-point equations $\partial_{b_i}S=0$ read
\eqn\sapno{V'(b_i)= {2\over N}\sum_{j\neq i}{1\over b_i-b_j}+{gn\over N}\sum_{j\neq i}
{1\over 1-g(b_i+b_j)} }
for $i=1,2,...,N$. Setting $x_i=1-2gb_i$, and $v'(x)={1\over 2g}V'(b)$, we get
\eqn\simpon{ v'(x_i)= {2\over N}\sum_{j\neq i}{1\over x_i-x_j}+{n\over N}\sum_{j\neq i}
{1\over  x_i+x_j} }
Multiplying this by ${1\over N}({1\over z-x_i}-{1\over z+x_i})$ and summing over $i$, we arrive
at the large $N$ quadratic equation:
\eqn\larnqa{\eqalign{ 
-Q(z)+v'(z)&\omega(z)+v'(-z)\omega(-z)=\omega(z)^2+\omega(-z)^2+
n\omega(z)\omega(-z)\cr
\omega(z)&=\lim_{N\to\infty}{1\over N}\sum_{i=1}^N {1\over z-x_i}\cr
Q(z)&=\lim_{N\to\infty}{1\over N}\sum_{i=1}^N 
{v'(z)-v'(x_i)\over z-x_i}-{v'(-z)-v'(x_i)\over z+x_i}\cr}}
while the saddle-point equation turns into a discontinuity equation across the support
of the limiting eigenvalue distribution $\rho$:
\eqn\discount{ \omega(z+i0)+\omega(z-i0)+n\omega(-z)=v'(z) \ \ z \in {\rm supp}(\rho) }
For generic $n$, $\omega(z)$ is well defined only in an infinite covering of the complex plane,
the transition from one sheet to another being governed by \discount. However, setting
\eqn\setneq{ n=2\cos(\pi \nu) }
we easily see that if $\nu=r/s$ is rational, this number becomes finite. Indeed, when expressed
in terms of the shifted resolvent
\eqn\shires{ w(z)={nv'(-z)-2 v'(z)\over n^2-4}+\omega(z) }
the discontinuity and quadratic equations \discount\ and \larnqa\ become respectively 
\eqn\newdisc{\eqalign{ w(z+i0)+w(z-i0)+nw(-z)&=0\cr
w(z)^2+w(-z)^2+nw(z)w(-z)&=P(z) \cr}}
where 
\eqn\eqP{ P(z)= {1\over n^2-4}(n v'(z)v'(-z)-v'(z)^2-v'(-z)^2)-Q(z) }
Introducing the two functions
\eqn\twofunc{ w_{\pm}(z)=e^{\pm i\pi{\nu\over 2}} w(z)+e^{\mp  i\pi{\nu\over 2}} w(-z) }
namely with $w_-(z)=w_+(-z)$,
we immediately get from \newdisc\ that
\eqn\diss{
\eqalign{ 
w_+(z+i0)&=-e^{i\pi\nu} w_-(z-i0)\cr
w_-(z+i0)&=-e^{-i\pi\nu} w_+(z-i0)\cr
w_+(z) w_-(z)&= P(z) \cr} }
For rational $\nu=r/s$, we see that the combination
\eqn\combi{ w_+(z)^s+(-1)^{r+s} w_-(z)^s= 2 S(z) }
is regular across the support of $\rho$. For instance, if $v(x)$ is polynomial (resp. meromorphic), 
so are $P(z)$ and $S(z)$. We may now solve \diss\-\combi\ for
$w_\pm$ as
\eqn\wpsol{ w_\pm (z)=(S(z)\pm\sqrt{S(z)^2-P(z)^s})^{1\over s} } 
and finally get the resolvent using \twofunc\-\shires:
\eqn\omp{\eqalign{
&\omega(z)={2 v'(z)-n v'(-z) \over n^2-4}- {1\over 2 i \sin(\pi {r\over s})} \times \cr
&\times (e^{i\pi{r\over 2s}}(S(z)+\sqrt{S(z)^2-P(z)^s})^{1\over s}
-e^{-i\pi{r\over 2s}}(S(z)-\sqrt{S(z)^2-P(z)^s})^{1\over s}) \cr}}
Assuming that $v$ is polynomial, we must further fix the coefficients of $P$ and $S$ by 
imposing the asymptotic behavior $\omega(z)\sim 1/z$ for large $|z|$. 
If deg$(v)=k$, then deg$(P)=2(k-1)$ and deg$(S)=s(k-1)$, and 
$P$ and $S$ are entirely fixed if we impose moreover that $\rho$ has a support made of a single 
interval $[a,b]$.

\subsec{Critical Behavior and Asymptotic Enumeration}

We have seen in Sect. 2 how to write the generating functions for various families
of  (decorated) fatgraphs in terms of Hermitian matrix integrals. It is a standard
fact that the critical properties of these generating functions can be translated into
asymptotics for these numbers of fatgraphs, say for large numbers of vertices. 
Indeed, writing such a generating function as $F(g)=\sum_{n\geq 0} F_n g^n$, where $F_n$
denotes a number of connected (decorated) fatgraphs with $n$ vertices,
assume $F$ has a critical singular part of the form
\eqn\critising{ F(g)_{sing}\sim (g_c-g)^{2-\gamma} }
when $g$ approaches some critical value $g_c$, then for large $n$ the numbers $F_n$ behave as
\eqn\fnbeha{ F_n \sim g_c^{-n} n^{-3+\gamma} } 

Let us apply this to the example of \onmav\ with the quartic potential 
$V(x)={x^2\over 2}-g{x^4\over 4}$. The generating function $F_N(g)={\rm Log}\, Z_N(V)=
\sum_{h\geq 0} N^{2-2h} f_{h}(g)$ decomposes into the generating functions $F_h(g)$
\eqn\genegg{ f_h(g)=\sum_{n\geq 0} g^n f_{h,n} }  
where $f_{h,n}$ denotes the number of fatgraphs of genus $h$ with $n$ vertices.  
Using the genus zero answer for $f_0(g)$ \genefg, we find that the critical singularity
is attained when $g=g_c={1\over 12}$, in which case $f_0(g)_{sing}\sim (g_c-g)^{5/2}$
and $\gamma=-1/2$ in \critising.
This implies the following asymptotics for the numbers of connected genus zero fatgraphs
\eqn\asymfg{ f_{0,n}\sim {{12}^n\over n^{7\over 2}} }
More refined applications of the orthogonal polynomial techniques lead to the genus $h$
result
\eqn\asymfgh{ f_{h,n}\sim {12^n \over n^{1+{5\over 2}(1-h)} } }
and to a score of interesting behaviors for the numbers of decorated fatgraphs
for specific potentials $V$, as well as for the O(n) model with also specific potentials,
for which any positive rational value of $-\gamma$ in \critising\ may be reached.
The exponent $\gamma$ is called the string susceptibility exponent. More generally,
it has been shown that the coupling of a critical matter theory to gravity results
in a behavior \critising\ of the genus zero free energy
with $\gamma$ given by the formula \KPZ\
\eqn\kpzone{ \gamma={c-1-\sqrt{(25-c)(1-c)} \over 12}}
where $c$ is the central charge of the conformal field theory underlying
the flat space critical matter model \CFT. The exponent $\gamma=-1/2$ is characteristic
of the ``pure gravity" models, in which there is no matter theory,
in other words $c=0$.
In the case of the so-called dense critical phase of the $O(n)$ model, one has \NIN\
\eqn\con{ c(n)= 1-6{e^2 \over 1-e} \ , \qquad n=2\cos(\pi e)}
and therefore
\eqn\gamon{ \gamma(n)=-{e\over 1-e} }
from \kpzone. This is confirmed by the saddle-point results \ONSOL\ mentioned above.

\subsec{Gaussian Words}

In the case of multi-matrix integrals, one may wonder how the beautifully simple
result \magau\-\mom\ is generalized. More precisely, we are interested in
the large $N$ limit of the multi-Gaussian average of the trace of any word:
\eqn\mutgau{ \eqalign{
&\eta_{n_1,n_2,...,n_{mk}}=
\langle\langle {\rm Tr}(M_1^{n_1}...M_k^{n_k}M_1^{n_{k+1}}...M_k^{n_{mk}}) 
\rangle \rangle\cr
&\equiv\lim_{N\to \infty}  
{\int dM_1...dM_k e^{-N{\rm Tr}(\sum_{i=1}^kM_i^2)}  
{\rm Tr}(M_1^{n_1}...M_k^{n_k}M_1^{n_{k+1}}...M_k^{n_{mk}}) \over
N\ \int dM_1...dM_k e^{-N{\rm Tr}(\sum_{i=1}^kM_i^2)} } \cr}}
In this setting, the result \magau\ corresponds to $k=m=1$ and reads 
\eqn\resone{\eta_{n}=\left\{\matrix{c_p  & {\rm if}\ n=2p\cr
0 & {\rm otherwise} \cr} \right. }
for all $n\geq 0$.
The simplest way of computing \mutgau\ in a given situation is the blind application of
Wick's theorem \muwic, by further restricting the Wick pairings to the 
{\it planar} ones only, as $N\to \infty$. For instance, for $k=2$ and $m=2$, we have
\eqn\wiap{\eqalign{ \eta_{n_1,n_2,n_3,n_4}&=
\langle\langle {\rm Tr}(M_1^{n_1}M_2^{n_2}M_1^{n_3}
M_2^{n_4}) \rangle\rangle \cr
&=\eta_{n_1+n_3}\eta_{n_2}\eta_{n_4}+\eta_{n_2+n_4}\eta_{n_1}\eta_{n_3}
-\eta_{n_1}\eta_{n_2}\eta_{n_3}\eta_{n_4}\ \cr}}
where we have isolated respectively the pairings in which elements of
the blocks
$M_1^{n_1}$ and $M_1^{n_3}$ may be connected, then those in which elements
of the blocks $M_2^{n_2}$ and $M_2^{n_4}$ may be connected, and we have subtracted 
the terms in which all pairings take place within the same powers of the $M$'s
to avoid counting them twice (once in each of the previous terms).  

To compute the most general trace of a word in say $k$ Gaussian Hermitian matrices, 
we simply have to apply the following quadratic recursion relation, for $\omega=e^{2i\pi/k}$
\eqn\recugauw{ 
\eta_{n_1,...,n_{mk}}=-\sum_{i=1}^{mk-1} \omega^i \eta_{n_1,...,n_i}
\eta_{n_{i+1},...,n_{mk}} }
This relation is a compact rephrasing of the Wick theorem in the case of planar pairings,
as the reader will check easily (the only relation to be used is 
$\sum_{0\leq i\leq k-1} \omega^i=0$, see \DGG\ for details).
This gives a priori access to the average of any trace of word.
The numbers $\eta$ are natural multi-dimensional generalizations of the Catalan numbers \catalan.

\vfill\eject
\centerline{\bf PART B: Polymer Folding and Meanders}
\vskip 1.truecm

We now present a first important application of matrix integrals to the fundamental
combinatorial problem of meander enumeration, also equivalent
to the compact folding problem of polymers. After defining meanders and reviewing
various approaches to their enumeration, we define a matrix model for meanders
and discuss its solution in some particular cases.
We finally present a general argument leading to the exact values of the 
meandric asymptotic configuration exponents.

\newsec{Folding Polymers: Meanders}

\subsec{Definitions and Generalities}

\fig{A typical polymer with $2n=10$ segments is depicted, together with 
one of its compact folding configurations, in which the segments have been 
deformed and slightly pulled apart for clarity. The corresponding meander
is obtained by drawing an horizontal line (river) intersecting the 
folding configuration (road) through $2n=10$ points (bridges). The connected half
segments of polymer have been replaced by semi-circular
portions of road for simplicity. }{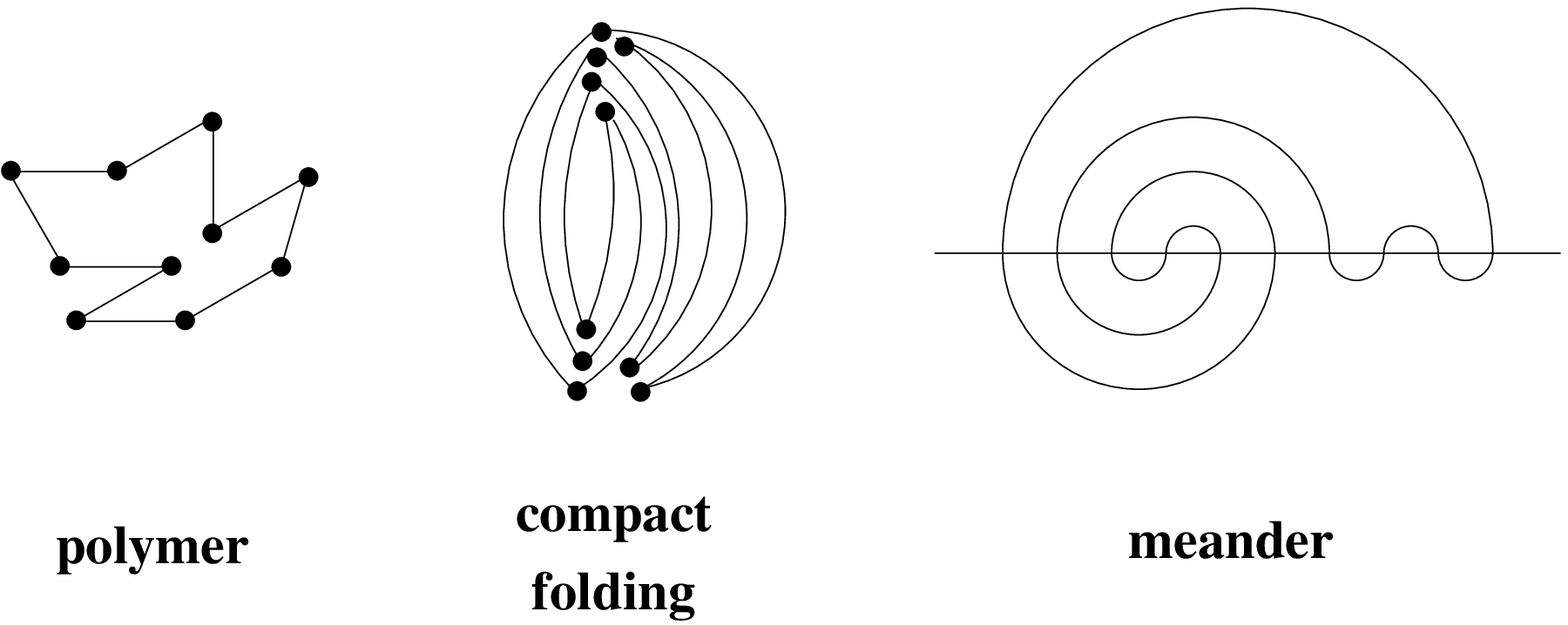}{10.cm}
\figlabel\folmean

A polymer is a chain of say $m$ identical constituents, modelled by rigid segments
attached by their ends, that serve as hinges in the folding process. A useful
representation to bear in mind is that of a strip of stamps of size $1\times m$,
attached by their ends. Given a closed polymer chain with $m=2n$ segments (i.e.
in which the chain is closed so as to form a loop, see Fig.\folmean), 
we address the question
of enumeration of all the ways of compactly folding the chain, in such a
way that in the final (folded) state all the segements are piled up onto one another.

As illustrated in Fig.\folmean, we may keep track of the folding by 
intersecting  the final state with an oriented line (say east to west), 
and slightly pulling apart the segments
of the folded polymer. 
Redrawing the result with a horizontal east-west oriented line,
and representing any two connected half-segments by semi-circles, we arrive at
a meander, i.e. the configuration of a non-intersecting loop (road, made of
semi-circular bits) crossing the horizontal line (river, flowing east to west) 
through a given number $m=2n$ of points (bridges), as depicted in Fig.\folmean.  
The number of bridges will also be called the order of the meander. The 
number of distinct meanders of order $2n$ is denoted by $M_{2n}$ \LZ.

We extend this definition to a set of $k$ roads namely meanders
with $k$ possibly interlocking connected components  and a total number of
bridges $2n$ (corresponding to the
simultaneous folding of $k$ possibly interlocking polymers with a total
number of $2n$ segments).
The number of meanders of order $2n$ with $k$ connected components
is denoted by $M_{2n}^{(k)}$. Note that necessarily $1\leq k \leq n$.
These numbers are summarized in the
meander polynomial \DGG\
\eqn\mepo{ m_{2n}(q)~=~ \sum_{k=1}^n M_{2n}^{(k)}~q^k  }
Given a meander of order $2n$, the river cuts the plane into two parts.
The upper part is called an arch configuration $a$ of order $2n$: it is
a set of $n$ non-intersecting semi-circles drawn in the upper
half plane delimited by the river and joining the $2n$ equally spaced bridges
by pairs. The lower part is clearly the reflection $b^t$ of another arch configuration
$b$ of order $2n$. Let $A_{2n}$ denote the set of arch configurations of order
$2n$, and for any $a,b \in A_{2n}$, let $c(a,b)$ denote the number of connected
components of road obtained by forming a (multi-component) meander with $a$ as
upper half and $b^t$ as lower half. Then we have
\eqn\pome{ m_{2n}(q)~=~
\sum_{a,b\in A_{2n}} q^{c(a,b)} }
The number of distinct arch configurations is nothing but the catalan number 
\eqn\catarch{ |A_{2n}|=c_n={(2n)!\over (n+1)! n!} }
as $A_{2n}$ can be easily mapped onto the planar Wick pairings of a single star with
$2n$ branches leading to \magau: indeed, we just have to pull all the bridges together
so as to form a $2n$-legged vertex; 
this results in a planar petal diagram like that of Fig.\planar\ (a). 
As an immediate consequence we get $m_{2n}(1)=|A_{2n}|^2=c_n^2\sim 4^{2n}/(\pi n^{3})$
by use of Stirling's formula for large $n$. In general, the meander polynomial is expected 
to behave for large $n$ as
\eqn\asympto{ m_{2n}(q)\sim c(q) {R(q)^{2n} \over n^{\alpha(q)}}}
and we just proved that $R(1)=4,\ \alpha(1)=3$ and $c(1)=1/\pi$. 
In Sect. 6.4 below, we will present an argument leading to the general 
determination of the exact value of the meander
configuration exponent $\alpha(q)$ as a  function of $q$. 
Note that 
\eqn\entme{s(q)=\lim_{n\to \infty} {1\over 2n}{\rm Log}\, m_{2n}(q)
={\rm Log}\, R(q)}
is nothing but the thermodynamic entropy of folding per segment of a multi-component
closed polymer.

\fig{An open polymer with $8$ segments, together with one typical compactly folded
configuration, and its semi-meander version of order $9$, obtained by intersecting 
the $8$ segments of the configuration with a circle (represented in dashed line),
that also intersects the support (wall) to which the open polymer is attached. The
free end of the folded polymer is then stretched into a straight half-line,
that becomes the semi-meander's river, the free end becoming its source. The dashed
circle is deformed into a winding road, crossing the river through $9$ bridges. 
For simplicity, we have represented all the pieces of the road joining the various
segments by means of semi-circles. Note that here the road winds three times
around the source of the river.}{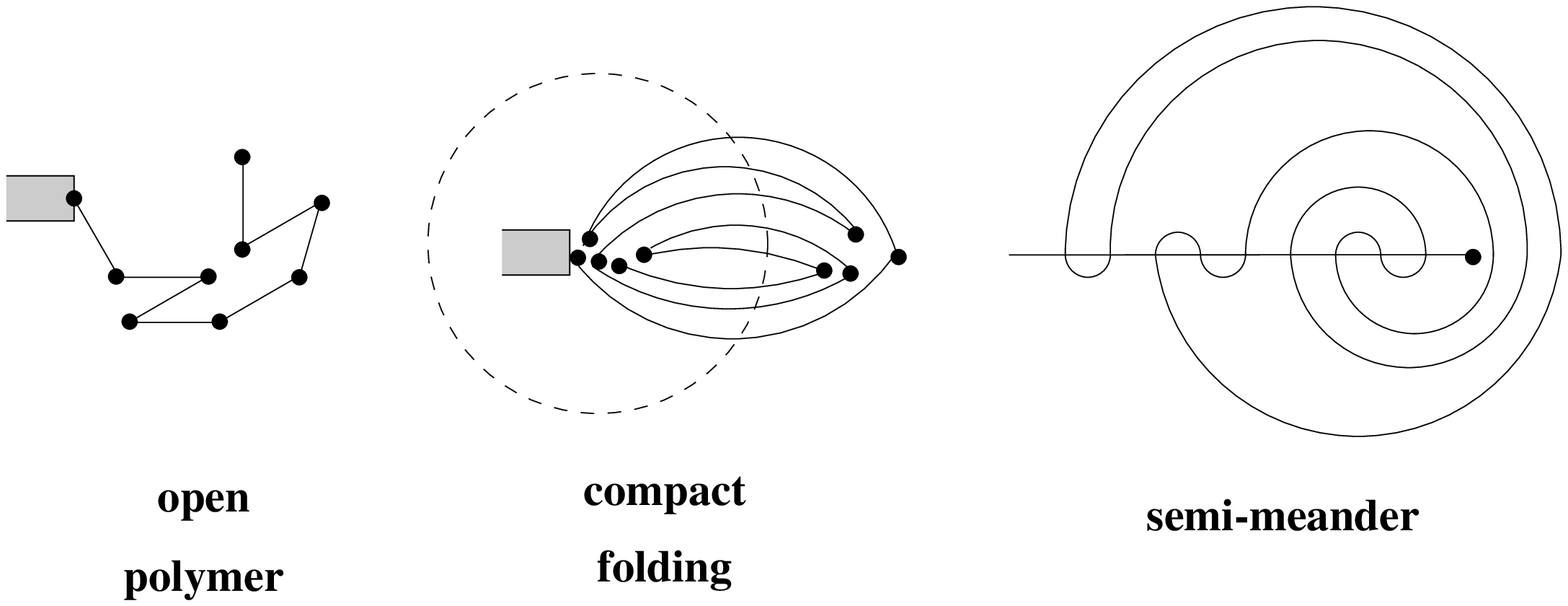}{11.cm}
\figlabel\folsem

We may also address the problem of compactly folding an open polymer chain of $n-1$
segments \DGG, attached to a wall by one of its ends (think of a strip of
stamps, attached in a book by its left end). 
As illustrated in Fig.\folsem, we may intersect the folded polymer with a
circle that also crosses the book's end. Extending the polymer itself
into a half-line, and deforming the circle accordingly without creating new
intersections, we arrive at a semi-meander of order $n$, i.e. the configuration 
of a closed non-self-intersecting loop (road, the former circle) crossing a 
semi-infinite line
(river with a source, the former polymer$+$book) through $n$ points (bridges).

Note that, in a semi-meander, the road may wind around the
source of the river, as illustrated in Fig.\folsem.
We denote by ${\bar M}_n$ the number of
topologically inequivalent semi-meanders of order $n$, and by
${\bar M}_n^{(k)}$ the number of semi-meanders with $k$ connected
components, $1 \leq k \leq n$. We also have the semi-meander polynomial
\eqn\sempo{{\bar m}_n(q)~=~ \sum_{k=1}^n {\bar M}_n^{(k)}~q^k }

\fig{A semi-meander of order $n$ is opened into a meander of order $2n$.
Think of the lower part of the river as pivoting around its source to form
a straight line, while all bridge connections are deformed without any new
intersections. The lower part of the meander is a rainbow, made of $n$ concentric
semi-circles. The winding number is just the number of semi-circular arches
passing at the vertical of the center (former source).}{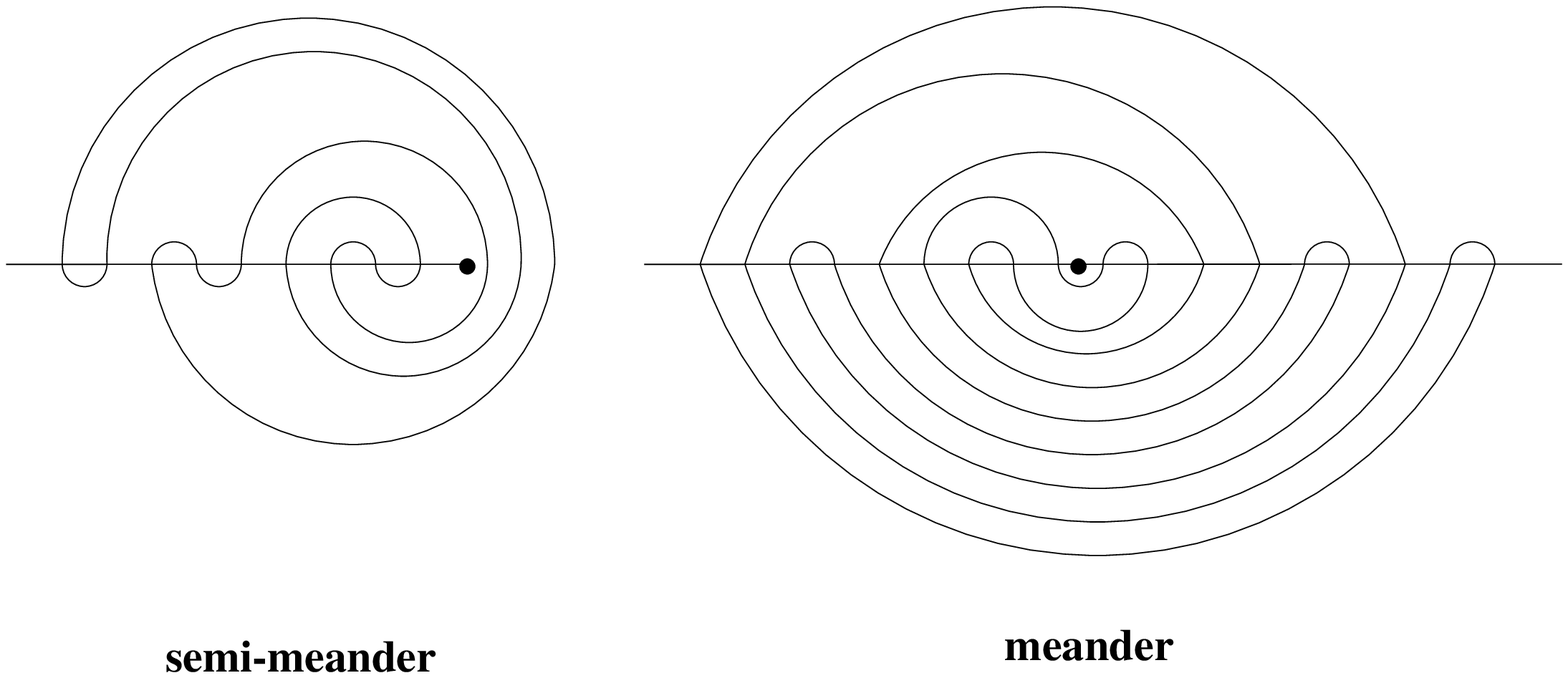}{10.cm}
\figlabel\openme

A semi-meander may be viewed as a particular kind of meander by opening the river
as sketched in Fig.\openme\ so as to double the number of bridges $n\to 2n$, and 
by connecting them as they were in the semi-meander in the upper-half of the
meander, and through the reflection of a rainbow arch configuration $r_{2n}^t$
in the lower one, made of $n$ concentric semi-circles.
The semi-meander polynomial is easily rewritten as
\eqn\semwr{ {\bar m}_n(q)=\sum_{a\in A_{2n}} q^{c(a,r_{2n})} }
and we have the value at $q=1$: ${\bar m}_{2n}(1)=c_n\sim 4^n/(\sqrt{\pi} n^{3/2})$.
We also expect the asymptotics
\eqn\asysem{ {\bar m}_{2n}(q)\sim {\bar c}(q) {{\bar R}(q)^n \over n^{{\bar \alpha}(q)}}}
and we have ${\bar R}(1)=4$, ${\bar\alpha}(1)=3/2$ and ${\bar c}(q)=1/\sqrt{\pi}$.
Note again that the semi-meander configuration exponent
${\bar \alpha}(q)$ will be determined as a function of $q$
in Sect. 6.4 below.

Conversely, a meander may be viewed as a semi-meander with no winding. It is therefore
sufficient to solve the semi-meander enumeration problem, provided
we keep track of the winding numbers.

\subsec{A Simple Algorithm; Numerical Results}

Multi-component semi-meanders of order $n$ are in one-to-one correspondence with 
arch configurations of order $2n$ \semwr. 
Our algorithm \NOUS\ is based on a recursive construction of arch configurations
that allows to keep track of both the numbers of connected components and windings.
To build an arch configuration of order $2n+2$ from one of order $2n$, we may 
do either of the following transformations, both involving the addition of two bridges
along the river, respectively to the left and right of the previous ones:

\item{(I)} For each external arch (semi-circle contained in no other) connecting
say the bridges $i$ and $j$, replace it by two semi-circles, one connecting the new
leftmost bridge to $i$, and one connecting $j$ to the new rightmost bridge.

\item{(II)} Add a large external arch connecting the two added bridges: it circles the 
whole previous arch configuration.
\par

\fig{The tree of semi-meanders down to order $n=4$.
This tree is constructed by repeated
applications of the transformations (I) and (II) on the semi-meander of order $1$
(root). We have indicated by small vertical arrows the multiple
choices for the process (I), each of which is indexed by its number.
The number of connected components of a given semi-meander is equal
to the number of processes (II) in the path going from the root
to it, plus one (that
of the root).}{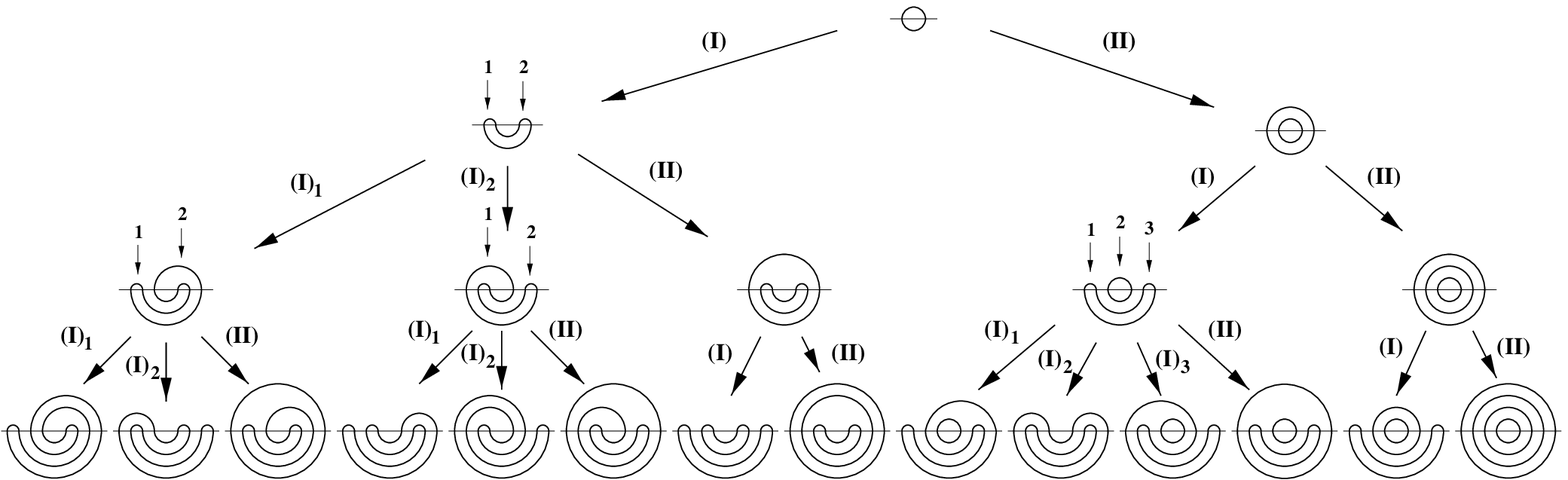}{12.cm}
\figlabel\treeofsm

The corresponding semi-meanders are obtained by completing this arch with the reflection
of the rainbow $r_{2n+2}$ as lower part. It is easy to show that applying 
$(I)-(II)$ to all of $A_{2n}$
yields exactly $A_{2n+2}$. Moreover, $(I)$ preserves the number of connected components,
while $(II)$ obviously increases it by $1$ (the net result is to add a circle around the
semi-meander). Applying successions of $(I)-(II)$ on the ``root" (semi-meander of
order $1$), we may build the tree of semi-meanders of Fig.\treeofsm.   

The above algorithm is easily implemented on a computer and yields 
all semi-meander numbers with fixed winding and connected components up to
some quite large orders. We give examples in Table I below.

$$\vbox{\font\bidon=cmr8 \bidon
\offinterlineskip
\halign{\tv \quad \hfill # & \hfill \ # \tv
& \quad \hfill # &  \hfill # \tv \tv
& \quad \hfill # &  \hfill # \tv
& \quad \hfill # &  \hfill # \tv \cr
\noalign{\hrule}
$n$ & $\bar M_n$ & $n$ & $\bar M_n$ & $k$ & $\bar M_{27}^{(k)}$
& $k$ & $\bar M_{27}^{(k)}$\cr
\noalign{\hrule}
1  &  1  & 16  &  1053874  &  1  &  369192702554  & 16  &  2376167414\cr
2  &  1  & 17  &  3328188  &   2  &  2266436498400 & 17  &  628492938\cr
3  &  2  & 18  &  10274466  &   3  &  6454265995454 &  18  &  153966062\cr
4  &  4  & 19  &  32786630  &   4  &  11409453277272 &  19  &  34735627\cr
5  &  10  & 20  &  102511418  &  5  &  14161346139866  &  20  &  7159268\cr
6  &  24  & 21  &  329903058  &  6  &  13266154255196 &  21  &  1333214\cr
7  &  66  & 22  &  1042277722  &  7  &  9870806627980 &  22  &  220892\cr
8  &  174  & 23  &  3377919260  &  8  &  6074897248976 &  23  &  31851\cr
9  &  504  & 24  &  10765024432  &  9  &  3199508682588 &  24  &  3866\cr
10  &  1406  & 25  &  35095839848  &  10  &  1483533803900 &  25  &  374\cr
11  &  4210  & 26  &  112670468128  &  11  &  619231827340  &  26  &  26\cr
12  &  12198  & 27  &  369192702554  &  12  &  236416286832 &  27  &  1\cr
13  &  37378  & 28  &  1192724674590  &  13  &  83407238044 &  &  \cr
14  &  111278  & 29  &  3925446804750  &  14  &  27346198448 &  &  \cr
15  &  346846  &     &                 &  15  &  8352021621 &  &  \cr
\noalign{\hrule}
}}$$
\noindent{\bf Table I:} The numbers $\bar M_n^{(k)}$ of semi-meanders
of order $n$ with $k$ connected components, obtained by exact
enumeration on the computer:  on the left, the one-component
semi-meander numbers ($k=1$) are given for $n \le 29$; on the right,
$n$ is fixed to 27 and $1 \leq k \leq n$.

These numbers allow in turn for probing the asymptotics \asympto\-\asysem, as functions of $q$.
In particular, we have the large $q$ asymptotics of the radii
\eqn\radis{\eqalign{
R(q)~&=~2 \sqrt{q} \big(1+{1\over q}+{3\over 2 q^2}-{3\over 2 q^3}
-{29\over 8 q^4}
-{81\over 8 q^5}-{89\over 16 q^6} +O({1\over q^7})\big) \cr
{\bar R}(q)~&=~ q+1+{2 \over q}+{2 \over q^2}
+{2 \over q^3}-{4 \over q^5}
-{8 \over q^6}-{12 \over q^7}-{10 \over q^8}-{4 \over q^9}
+{12 \over q^{10}}
+{46 \over q^{11}}\cr
&+{98 \over q^{12}}+{154 \over q^{13}}+{124 \over q^{14}}+{10\over q^{15}}
-{102 \over q^{16}}+{20 \over q^{17}}-{64 \over q^{18}}
+O({1 \over q^{19}})\cr }}
\fig{The functions ${\bar R}(q)$ and $R(q)$ for $0 \leq q\leq 4$
as results of large $n$ extrapolations. The two curves coincide
for $0 \leq q\leq 2$ and split for $q>2$
with ${\bar R}(q)>R(q)$. Apart from the exact value ${\bar R}(1)=R(1)=4$,
we find the estimates ${\bar R}(0)=3.50(1)$, ${\bar R}(2)=4.44(1)$,
${\bar R}(3)=4.93(1)$ and ${\bar R}(4)=5.65(1)$.}{rbb.eps}{6. cm}
\figlabel\rbarbare

At finite values of $q$, the numerical results displayed in Fig.\rbarbare\
reveal an interesting phase transition
between a phase for $q<q_c$ of irrelevant winding, i.e. the numbers of meanders 
and semi-meanders are then
asymptotically equivalent $R(q)={\bar R}(q)$), and a strong winding phase $q>q_c$, 
where the winding number is proportional to $n$, and therefore ${\bar R}(q)>>R(q)$. 
The transition point is estimated as $q_c\simeq 2$ with poor precision. We will 
propose an exact value for $q_c$ in Sect. 6.4 below. 

\newsec{Algebraic Formulation: Temperley-Lieb Algebra}

\subsec{Definition}

The Temperley-Lieb algebra of order $n$ and parameter $q$, denoted by
$TL_n(q)$, is defined through its $n$ generators
$1,e_1,e_2,...,e_{n-1}$ subject to the relations
\eqn\tla{\eqalign{(i)\ \ \ \ \ \ \ \ \  e_i^2 ~&=~
q \, e_i \quad i=1,2,...,n-1\cr
(ii)\ \ \ \ [e_i,e_j]~&=~0 \quad {\rm if}\ |i-j|>1 \cr
(iii)\ e_i\, e_{i \pm 1}\, e_i~&=~ e_i  \quad i=1,2,...,n-1\cr}}
This definition becomes clear in the ``domino" pictorial representation,
where the generators  are represented as dominos as follows:
\eqn\braid{\eqalign{ 1~&=~\figbox{2.cm}{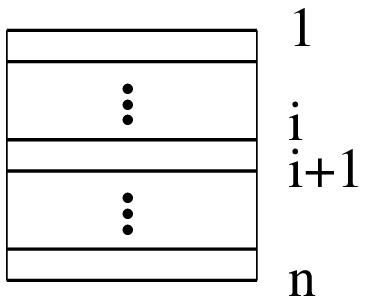} \cr
e_i~&=~\figbox{2.cm}{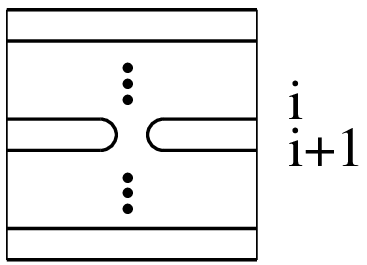}\cr} }
and a product of elements is represented by the concatenation of the
corresponding dominos. Note that we have numbered the left
and right ``string ends" of the dominos
from top to bottom, $1$ to $n$.
The relation (ii) expresses the locality of the $e$'s,
namely that the $e$'s commute whenever they involve distant strings.
The relations (i) and (iii) read respectively
\eqn\unbraid{\eqalign{(i)\ \ \ \ \ \ \ \ \ e_i^2~&=~
\figbox{2cm}{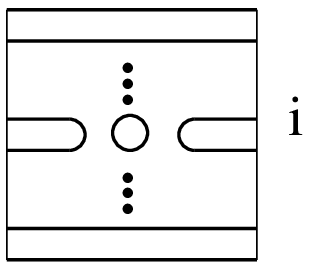}\cr
&=~q \, e_i~=~q~\figbox{2.4cm}{ei.eps}\cr
(iii)\ e_i\, e_{i+1}\, e_i~&=~
\figbox{2.4cm}{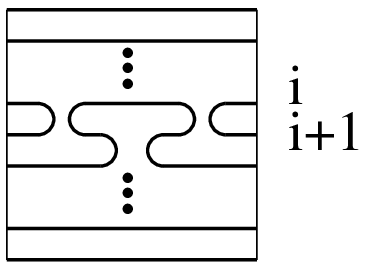}\cr
&=~e_i~=~\figbox{2.4cm}{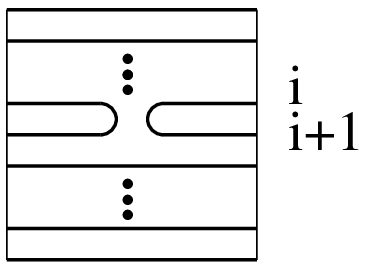}\cr}}
In the relation (i), the loop has been erased, but affected the weight $q$.
The relation (iii) is simply obtained by stretching the $(i+2)$-th string.

Moreover the algebra is equipped with a trace, defined as
follows. Given a domino $d$, we put it on a cylinder by identifying
its left and right string ends. We then count the number $n(d)$ of
connected strings in the resulting picture. The trace is then
simply ${\rm Tr}(d)=q^{n(d)}$. This definition is extended to any
element of the algebra by linearity.
This trace has the important Markov property which allows to 
compute it by induction:
\eqn\marko{ {\rm Tr}(d(e_1,e_2,...,e_j) e_{j+1})={1\over q}
{\rm Tr}(d(e_1,e_2,...,e_j))}
for any algebra element $d(e_1,...,e_j)$ involving only the generators
$e_1,e_2,...e_j$.

The crucial remark actually concerns the left ideal $I_{2n}(q)$
of $TL_{2n}(q)$ generated by the element $e_1e_3...e_{2n-1}$. The
corresponding dominos have their right string ends paired by single
arches linking the ends $2i-1$ and $2i$, $i=1,2,...,n$. Therefore
their $2n$ left string ends are also connected among themselves,
thus forming an arch configuration of order $2n$ after eliminating
the internal loops and stretching
the strings  (by use of $(i)-(iii)$) and giving them a semi-circular shape.
We therefore have a bijection between  the set $A_{2n}$ of arch 
configurations of
order $2n$ and a basis of the ideal $I_{2n}(q)=TL_{2n}(q) e_1e_3...e_{2n-1}$,
made only of ``reduced" dominos, namely with all their loops removed and their
strings stretched. The set of reduced dominos of  $I_{2n}(q)$
is dnoted by $D_{2n}$. As a consequence we have dim$(I_{2n}(q))=c_n$.
We denote by the same letter $a\in A_{2n}$ or $D_{2n}$ 
the arch configuration or the corresponding reduced domino.

\subsec{Meander Polynomials}

Given a pair $a,b$ of reduced dominos of $I_{2n}(q)$, we may form 
the scalar product
\eqn\scatla{ (a,b)= {1\over q^n} {\rm Tr} (b^t a) }
where by $b^t$ we mean the reflected domino wrt one of its vertical
edges. The middle of the concatenated domino $b^ta$ is nothing
but the meander with $a$ as upper-half and $b^t$ as lower one
(tilted by $90^\circ$), while it also contains $n$ small circles
formed in the cylinder identification, hence
\eqn\tramed{ (a,b) = q^{c(a,b)} }
We deduce the following expressions for the meander and semi-meander 
polynomials in purely algebraic terms \NTLA:
\eqn\alterms{\eqalign{
m_{2n}(q)&= \sum_{a,b\in D_{2n}} (a,b) \cr
{\bar m}_n(q)&=\sum_{a\in D_{2n}} (a,r_{2n})\cr}}
As the trace can be computed by induction, using \marko, we now
get an inductive way of computing the meander and semi-meander polynomials.
This is however far from explicit, and does not allow a priori for a
good asymptotic study of these polynomials.

\subsec{Meander Determinants}

On the other hand, we may consider other meander-related quantities
which are explicitly calculable by use of the Temperley-Lieb algebra 
representation theory. the most interesting of them is the 
``meander determinant" constructed as follows. We first define
the Gram matrix for the reduced basis of $I_{2n}(q)$ as
the $c_n\times c_n$ matrix $G_{2n}(q)$ with entries
\eqn\gram{ G_{2n}(q)_{a,b} = (a,b) \ \ \ \ \ \forall \ a,b \in
D_{2n} }
The meander determinant is then defined as the Gram determinant
\eqn\gramdet{ \Delta_{2n}(q)= \det(G_{2n}(q)) }

This determinant is computed by performing the explicit Gram-Schmidt
orthonormalization of the scalar product $(.,.)$, and using along
the way the representation theory of $TL_n(q)$. The result is
remarkably simple \NTLA\-\MEDET\
\eqn\resdet{\eqalign{
\Delta_{2n}(q)&= \prod_{m=1}^n U_m(q)^{a_{m,n}} \cr
a_{m,n}&= {2n \choose n-m} -2 {2n \choose n-m-1}
+{2n\choose n-m-2}\cr}}
where the $U_m(q)$ are the Chebyshev polynomials of the second kind
defined by $U_m(q)=\sin((m+1)\theta)/\sin(\theta)$ with $2\cos \theta=q$.

\newsec{Matrix Model for Meanders}

We have seen in Sect. 2 how to tailor matrix integrals to suit our combinatorial
needs. The meanders are basically obtained by intersection of two types
of curves: the roads and the rivers. In the following, we will define a matrix
model that generates arbitrary configurations of any number of rivers (now viewed as
closed curves, i.e. in the case of just one river we should close it into a circle),
crossed by any number of non-(self)intersecting roads \DGG.

\subsec{The B$\&$W Model}

Let us now construct a Hermitian
matrix integral that generates the meander polynomials.
The computation of such an integral must involve fatgraphs
with double-line edges, which we will eventually interpret as
the river(s) and the road(s).
Let us paint in white the river edges, and in black the road
edges. We therefore have a ``black and white" graph made of black
and white loops which intersect each other through simple intersections.
To assign a weight say $q$ per black loop (component of road) and
$p$ per white loop (component of river), the simplest way is to use
the  replica trick of Sect. 2.3: for positive integer values of $p$ and $q$,
introduce $q$ ``black" Hermitian matrices
$B_1,B_2,...,B_q$
and $p$ ``white" Hermitian matrices $W_1,W_2,...,W_p$,
all of size $N\times N$, with the only non-vanishing
propagators
\eqn\bwprop{\eqalign{
{\rm white} \ {\rm edges}&:
\ \ \ \langle (W_a)_{ij} (W_b)_{kl} \rangle~=
{1 \over N} \delta_{a,b}\delta_{il}\delta_{jk}=\figbox{1.3cm}{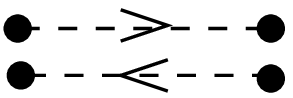} \cr
{\rm black}\ {\rm edges}&:
\ \ \ \langle (B_a)_{ij} (B_b)_{kl} \rangle ~=~
{1 \over N} \delta_{a,b}\delta_{il}\delta_{jk} =\figbox{1.3cm}{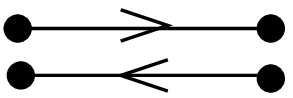}\cr}}
and only simple intersection vertices 
\eqn\intervert{{\rm Tr}(W_a B_b W_a B_b) =\figbox{2.cm}{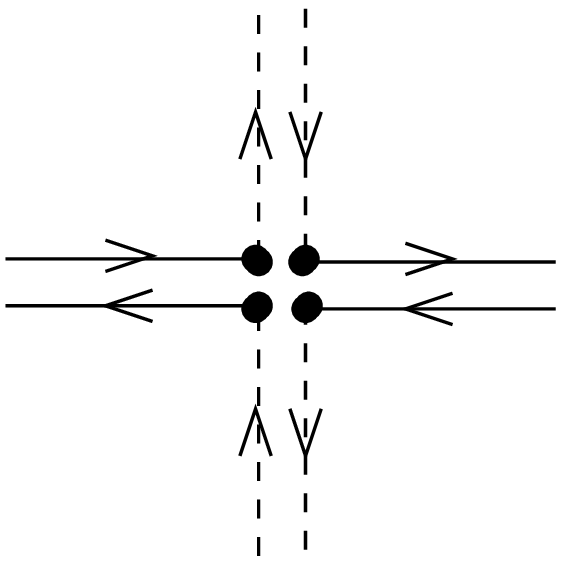} }
for all $1\leq x\leq p$, $1\leq y\leq q$.

The case of a unique river will then be recovered by taking the limit $p\to 0$.
This suggests to introduce the ``black and white" matrix integral
\eqn\bw{\eqalign{
Z_{q,p}(N;x)~&=~ {1\over Z_0}\int \prod_{a=1}^q dB_a \prod_{b=1}^p dW_b
e^{-N{\rm Tr}\, V(\{B_a\} ,\{ W_b\})}\cr
V(\{B_a\} ,\{ W_b\})~&=~ {1 \over 2}\big(\sum_{a=1}^q B_a^2 +\sum_{b=1}^p W_b^2
-x \sum_{a=1}^q \sum_{b=1}^p B_a W_b B_a W_b\big) \cr}}
where $Z_0$ is a normalization factor ensuring that $Z_{q,p}(N;0)=1$.
As before, the corresponding free energy may be formally expanded as a sum over
all possible connected black and white graphs as
\eqn\frebw{\eqalign{
F_{q,p}(N;x)~&=~{1 \over N^2} {\rm Log}\, Z_{q,p}(N;x) \cr
&=~ \sum_{{\rm black}\ \& {\rm white}\ \Gamma}
{1 \over |{\rm Aut}(\Gamma)|} N^{-2g(\Gamma)} x^{v(\Gamma)} q^{L_b(\Gamma)}
p^{L_w(\Gamma)} \cr}}
where, as in Sect. 2.2, $g(\Gamma),v(\Gamma)$ and Aut$(\Gamma)$
denote respectively the genus, number of vertices,
and symmetry group of $\Gamma$, whereas $L_b(\Gamma)$ and $L_w(\Gamma)$
denote respectively the numbers of black and white loops of resp. black and
white edges in $\Gamma$.
To get a generating function for meander polynomials from \frebw, we simply have
to take $N\to \infty$ to only retain the planar graphs (genus zero), and
then only compute the coefficient of $w$ in the resulting expression
as a power series of $w$, which leads to
\eqn\genemewb{F_q(x)\equiv  \lim_{N\to \infty} 
{\partial F_{q,p}(N;x)
\over \partial p}\bigg\vert_{p=0}~
=~ \sum_{n=1}^\infty {x^{2n} \over 4n}\,
m_{2n}(q) }
where we simply have identified planar black and white
fatgraphs with meanders (with the river closed into a loop), whose symmetry group is 
$\IZ_{2n}\times \IZ_2$ for the cyclic symmetry along the looplike river,
and the symmetry between inside and outside of that loop. The meander polynomial is
as in \mepo.
The large $n$ asymptotic behavior of the meander polynomial $m_{2n}(q)$ can be
directly linked to the critical behavior of the generating function $F_q(x)$ \genemewb.
Indeed, \asympto\ translates into a singular part 
\eqn\singF{ F_q(x)_{\rm sing}\sim (x(q)-x)^{\mu(q)} }
where 
\eqn\xandmu{ x(q)= {1\over R(q)} \qquad \mu(q)=\alpha(q)-1 }
So, if we can investigate the critical properties of $F_q(x)$, the meander
asymptotics will follow.

The B$\&$W model in the large $N$ limit
also allows for a simple representation of semi-meanders
as a correlation function. Indeed, considering the operator
\eqn\ophi{ \phi_1=\lim_{N\to \infty} {1\over N} {\rm Tr}(W_1) }
and computing the correlation function
\eqn\corela{
\langle \phi_1 \phi_1 \rangle_{q,p} =
{1\over Z_0} \int \prod dW_b dB_a \phi_1 \phi_1
e^{-N{\rm Tr}\, V(\{B_a\} ,\{ W_b\})} }
by use of the Wick theorem, we see that $\phi_1$ creates a white
line at a point, hence the fatgraphs contributing to the $p\to 0$ limit
will simply have a white segment joining the two endpoints created by
the two $\phi_1$'s, intersected by arbitrary configurations of road.
Given such a planar graph, we may always send one of the two endpoints 
to infinity thus yielding a semi-infinite river, and the fatgraphs with
$n$ intersections are just
the semi-meanders of order $n$. Hence
\eqn\semimatri{ \sum_{n\geq 1}{\bar m}_n(q)x^n= \langle \phi_1\phi_1 \rangle_{q,0}}
Again, we will get the semi-meander asymptotics from the singular behavior of this
correlation function when $x\to x_c$.

Before taking the $p\to 0$ limit, we could have written
\eqn\Ftotal{ F_{q,p}(x)\equiv \lim_{N\to \infty} F_{q,p}(N;x)=  
\sum_{n=1}^\infty {x^{2n} \over 4n}\,
m_{2n}(q,p) }
where we have defined a meander polynomial $m_{2n}(q,p)$ for multi-component meanders
with also multiple rivers (and a weight $q$ per road and $p$ per river). 
In particular $m_{2n}(q,p)\sim p m_{2n}(q)$ when $p\to 0$.
Similarly, 
\eqn\corsem{ \langle \phi_1\phi_1\rangle_{q,p}=\sum_{n\geq 1} {\bar m}_n(q,p) x^n}
defines a semi-meander polynomial with also multiple rivers, one of which
is semi-infinite.
In Sect. 6.3 below, we will derive the exact asymptotics of the polynomial
$m_{2n}(q,p=1)$ from the Black and White matrix model, while in Sect. 6.4, we will
present an argument yielding all the critical configuration exponents for large $n$.

Note also that if we keep $N$ finite, we get an all genus expansion
\eqn\allgebw{{\partial F_{q,p}(N;x)\over \partial p}\bigg\vert_{p=0}~=~
\sum_{g \geq 0} N^{-2g}  \sum_{n=1}^\infty {x^{n} \over 2n}\,
m_{n}^{(g)}(q) }
where we have defined the genus $g$ meander polynomials $m_{n}^{(g)}(q)$ that counts
the natural higher genus generalization of meanders of given genus and number of connected
components of road, and with one river.

\subsec{Meander Polynomials and Gaussian Words}

The integral \bw\ can be considerably simplified  by noting that it is just 
multi-Gaussian say in the white matrix vector $\vec{W}=(W_1,...,W_p)$. 
Using the recipees of Sect. 3.4, we first identify the quadratic form involving
each $W_b$ in \bw, namely
\eqn\qafow{\eqalign{ 
W_b^t {\bf Q} W_b &= \sum_{i,j,k,l=1}^N (W_b)_{ji} Q_{ij,kl}
(W_b)_{kl}\cr
&={\rm Tr}\big( W_b^2 -x \sum_{a=1}^q
W_bB_aW_bB_a \big) \cr}}
hence
\eqn\qaexp{
Q_{ij,kl}= \delta_{ik}\delta_{jl} -x  \sum_{a=1}^q
(B_a)_{ik} (B_a)_{lj} }
or more compactly
\eqn\compaq{ {\bf Q} = I \otimes I -x \sum_{a=1}^q B_a \otimes B_a^t}
After integration over the $W$'s in \bw,  
we are left with
\eqn\intewq{ Z_{q,p}(N;x)= {\int \prod_{a=1}^q dB_a
e^{-{N\over 2}{\rm Tr}B_a^2} \det({\bf Q})^{-p/2}\over \int 
\prod_{a=1}^q dB_ae^{-{N\over 2}{\rm Tr}B_a^2} }}
Extracting the $p\to 0$ limit is easy, with the result
\eqn\reslimip{  
{\partial F_{q,p}(N;x)
\over \partial p}\bigg\vert_{p=0}~=~ 
{1\over N^2}\langle -{1\over 2}{\rm Tr} \, {\rm Log}\, {\bf Q} \rangle }
where the bracket stands as in Sect. 2.3 for the multi-Gaussian average wrt the $B$'s.
Expanding Log $\bf Q$ of \compaq\ as a formal power series of $x$, we get 
finally
\eqn\larNft{ {\partial F_{q,p}(N;x)
\over \partial p}\bigg\vert_{p=0}~=~\sum_{m\geq 1} {(-x)^{m}\over 2m} 
\sum_{a_1,a_2...,a_m=1}^q \langle
\vert {1\over N}{\rm Tr}(B_{a_1}B_{a_2}...B_{a_m}) \vert^2 \rangle }
Comparing this with \allgebw, we identify the all-genus meander polynomial generating function 
\eqn\mepog{
\sum_{g \geq 0} N^{-2g} m_{n}^{(g)}(q)=\sum_{a_1,a_2...,a_n=1}^q \langle
\vert {1\over N}{\rm Tr}(B_{a_1}B_{a_2}...B_{a_n}) \vert^2 \rangle }

Concentrating on the genus zero case,
we may express the meander polynomial \mepo\ as a multi-Gaussian integral
\eqn\mulmep{ m_{2n}(q)=\sum_{a_1,a_2...,a_{2n}=1}^q \lim_{N\to \infty} \langle
\vert {1\over N}{\rm Tr}(B_{a_1}B_{a_2}...B_{a_n}) \vert^2 \rangle }
where we have noted that the rhs of \mepog\ vanishes for large $N$ if $n$ is odd, by a simple
parity argument. Moreover, the rhs of \mulmep\ must be computed using the planar
Wick theorem. It is clear however that no pairing can be made between $B$ matrix elements
pertaining to the two trace terms, as that would violate planarity. 
In other words, in the large $N$ limit, we have 
$\langle {\rm Tr}\, f(B_j) {\rm Tr}\, g(B_j)\rangle=\langle {\rm Tr}\, f(B_j)\rangle\langle
{\rm Tr}\, g(B_j)\rangle$ for any functions $f$ and $g$ of the $B$'s.
Hence, using
the definition \mutgau\ of Sect. 3.6, we may finally express the meander polynomial in
terms of planar multi-Gaussian averages of words
\eqn\mepogow{ m_{2n}(q)=\sum_{a_1,a_2...,a_{2n}=1}^q 
\vert \gamma_{a_1,a_2,...,a_{2n}}\vert^2 }
where we have set for $a_1,...a_p\in\{1,2,...,q\}$ 
\eqn\gamet{\eqalign{ \gamma_{a_1,a_2,...,a_p} &=\eta_{m_1,...,m_{qr-1}} \cr
{\rm if}
\ \ \ B_{a_1}B_{a_2}...B_{a_p}&= B_1^{m_1} ...B_q^{m_q}B_1^{m_{q+1}}...B_q^{m_{qr-1}}\cr}}

\fig{A typical pair of Wick pairings used to compute the quantity
$|\gamma_{1,2,2,1,3,3,1,1,1,1,3,3}|^2$. Each pair is represented by a semi-circle
with the corresponding color (1=solid, 2=dashed, 3=dotted). A pairing for $\gamma$
is represented, as a colored arch configuration of order 12 here. 
A pairing for 
$\gamma^*$ is also represented, but head-down, to match the bridge colors.
The net result is a colored meander with 3 components of road. }{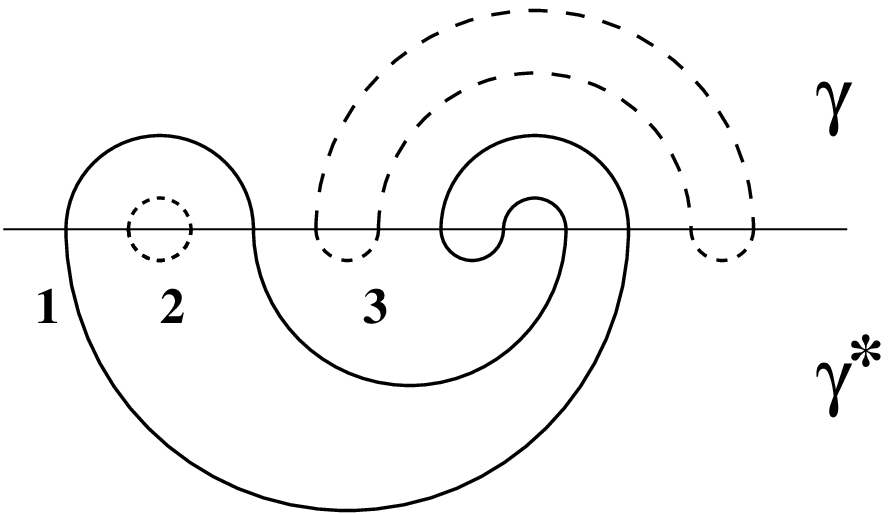}{6.cm}
\figlabel\wickmean

Having proved \mepogow\ through some quite lengthy process, it is instructive to
interpret \mepogow\ directly. Using the planar Wick theorem to compute the rhs of \mepogow,
let us represent for $\gamma_{a_1,...,a_{2n}}$
the chain of $2n$ matrix elements as equally spaced points along a line, each with the
corresponding  color $a_i\in \{1,2,...,q\}$ (see Fig.\wickmean\ for an
illustration). A given planar Wick pairing is nothing but 
an arch configuration linking these points by pairs, {\it provided they have the same color}.
As we have to multiply two $\gamma$'s, we get a pair of such colored arch configurations.
The fact that the two $\gamma$'s are conjugate of one another simply means that the two
arch configurations may be superposed, and that their bridge colors match: indeed,
$\gamma_{a_1,..a_{2n}}^*=\gamma_{a_{2n},...,a_1}$, so we just have to represent
its pairings head down and use the same line as for $\gamma$. So the net result
is the production, for each (pair of) Wick pairing pertaining to the two $\gamma$'s and
for each choice of colors $a_1,a_2,...,a_{2n}$ of the bridges, of a {\it colored} 
multi-component meander, namely with colored roads. As the colors are summed over, we get
a final weight of $q$ per connected component of road, which matches the definition
of the meander polynomial.

This interpretation of the result \mepogow\ allows to immediately generalize it to
the semi-meander case:
\eqn\semibw{ {\bar m}_n(q)= \sum_{a_1,a_2,...,a_n=1}^q 
\gamma_{a_1,a_2,...,a_n,a_n,a_{n-1},...,a_2,a_1} }
Indeed, the planar Wick pairings for the $\gamma$ form again colored arch configurations
of order $2n$ with the constraint that they have symmetrically identical colors. This 
latter constraint can be represented  by a lower rainbow diagram $r_{2n}$ with the
corresponding colors. The net result is a colored semi-meander, and the result \semibw\
follows exactly like before. 

By analogy with \mepog, we may also define the genus $g$ semi-meander  polynomials
as 
\eqn\semgpo{   
\sum_{g \geq 0} N^{-2g} {\bar m}_{n}^{(g)}(q)=\sum_{a_1,a_2...,a_n=1}^q \langle
\vert {1\over N}{\rm Tr}(B_{a_1}B_{a_2}...B_{a_n}B_{a_n}...B_{a_1}) \rangle }

Having expressed all our quantities of interest in terms of $\gamma$'a, we now have an
alternative route for evaluating them. We have to use the recursion relations 
\recugauw\ to first compute the $\gamma$'s and then to substitute them in the above
expressions for (semi-)meander polynomials.
This is still however a tedious work, and it doesn't really allow for any 
asymptotic study at large $n$. One advantage though of these expressions
is that as the $\gamma$'s are positive numbers, they allow for exact lower
bounds on the meander polynomials.

A last remark is in order: in this section, the meander and semi-meander polynomials
have been computed only for positive integer values of $q$. Analytic continuation to 
any positive real $q$ is however immediate as these are polynomials.

\subsec{Exact Asymptotics for the case of arbitrary many rivers}

In this section, following \CK, we will use the black and white matrix model for meanders to 
evaluate the exact asymptotics of slightly different quantities, namely 
the ``multi-meander" polynomials $m_{2n}(q,1)$ counting the meanders with fixed numbers of
connected components of roads and arbitrary numbers of rivers, closed into 
non-intersecting loops. However this side result already announces
the flavor of what true meander asymptotics should look like.

We wish to compute the function $Z_{q,1}$ \bw\ by first integrating over all the
$B$ matrices, rather than the single $W$ one. Using the form
\compaq\ with $B\to W$, we are then left with
\eqn\onew{
Z_{q,1}(N;x)= {\int dW e^{-N{\rm Tr}(W^2/2)} \det(1\otimes 1-x W\otimes W^t)^{-q/2} \over
\int dW e^{-N{\rm Tr}(W^2/2)} } }
This is just a Gaussian average over one Hermitian matrix $W$. 
Applying the reduction method of Sect. 3.1,
we arrive at the eigenvalue integral
\eqn\evawone{
Z_{q,1}(N;x)= {1\over (2\pi N)^{N/2}}\int dw_1 ... dw_N  {\prod_{i<j}(w_i-w_j)^2 \over
\prod_{i,j} (1-x w_iw_j)^{q/2} } 
e^{-N \sum_i {w_i^2\over 2}} }
This is very similar to the eigenvalue integral \eigint\ we obtained in the case of the 
O(n) model of Sect. 3.4. Actually, upon the change of variables 
\eqn\chvarwz{ w={1\over \sqrt{x}} {1-z\over 1+z} }
we have 
\eqn\after{\eqalign{
&Z_{q,1}(N;x)= {2^{{N^2\over2}(2-q)}
\over (2\pi N x)^{N/2}}\times \cr
&\times \int dz_1 ... dz_N {\prod_{i<j}(z_i-z_j)^2\over   
\prod_{i,j} (z_i+z_j)^{q/2}} \prod_{i} (1+z_i)^{N(q-2)} e^{-N\, \sum_i 
{1\over 2x}\big({1-z_i\over 1+z_i}\big)^2} \cr}}
Let us now evaluate the large $N$ behavior of the integral using the saddle-point technique
of Sect. 3.4. The action to be minimized reads now
\eqn\action{\eqalign{
S(z_1,...,z_N)&= {1\over N}\sum_{i=1}^N v(z_i) -{1\over N^2}\sum_{1\leq i\neq j\leq N}
{\rm Log}|z_i-z_j|
-{q\over 2N^2}\sum_{1\leq i,j\leq N} {\rm Log}(z_i+z_j) \cr  
v(z)&={1\over 2x}\big({1-z\over 1+z}\big)^2 +(q-2){\rm Log}(1+z) \cr}}
Note that the sum in the last term of $S$ may be restricted to $i\neq j$ up to a term of
order $O(1/N)$ that does not contribute to the large $N$ leading asymptotics.
Expressing that $\partial_{z_i} S=0$, we finally get
\eqn\sapobw{
v'(z_i)= {2\over N} \sum_{j\neq i} {1\over z_i-z_j} +{q\over N}\sum_{j\neq i} {1\over z_i+z_j}}
This is exactly the saddle-point equation \simpon\ for $n=q$ and the  particular choice 
\action\ of the potential $v(z)$. Note that $v'$ is a meromorphic function of $z$ with a
third order pole at $z=-1$. As before, we assume that the limiting eigenvalue distribution
has a support made of a single interval $[a,b]$, $0<a<b$. 
This requirement turns out to fix entirely the meromorphic functions $S(z)$ and $P(z)$
in \omp. The critical singularity of the genus zero free energy $f\sim (x(q)-x)^{2-\gamma}$
is found to lie
at \CK\
\eqn\critix{ x=x(q)={f^2 \over 2 \sin(\pi {f \over 2}) } }
where we have set $q=2 \cos(\pi f)$, and the corresponding critical exponent $\gamma$
reads
\eqn\gavalon{ \gamma=-{f \over 1 - f} }
As mentioned before, these translate into multi-river meander asymptotics 
\eqn\asymenla{m_{2n}(q,1)\sim {R(q,1)^{2n}\over n^{\alpha(q,1)} }}
as
\eqn\randalph{\eqalign{
R(q,1)&={1\over x(q)}=2 {\sin^2(\pi {f\over 2}) \over f^2} \cr
\alpha(q,1)&={2-f \over 1-f} \cr}}
In particular, we recover from these values the case of meanders with one river and arbitrary
many connected components of road, with $q=0$, $f=1/2$, and $R(1,0)=R(0,1)=4$,
$\alpha(1,0)=\alpha(0,1)=3$. We list a few of these values for various fractions $f$
in Table II below.

$$\vbox{\font\bidon=cmr8 \bidon
\offinterlineskip
\halign{\tv \quad \hfill # \hfill &\tv \hfill  \ # \hfill &\tv
\hfill # \hfill &\tv  \hfill #  
\hfill &\tv  \hfill # \hfill \tv \cr
\noalign{\hrule}
$q$ & $f$ & $\ \ \ R(q,1)\ \ \ $ & $\ \ \ \alpha(q,1)\ \ \ $ & $\ \ \ R(q)\ \ \ $ \cr
\noalign{\hrule}
0  &  ${1\over 2}$  & 4  &  3  &  3.50  \cr
1  & ${1\over 3}$   & ${9\over 2}$  & ${5\over 2}$   & 4    \cr
$\sqrt{2}$  & ${1\over 4}$   & $16-8\sqrt{2}=4.68...$ & ${7\over 3}$  & 4.13...   \cr
$\sqrt{3}$  & $ {1\over 6}$  & $36-18\sqrt{3}=4.82...$  & ${11\over 5}$  & 4.27...  \cr
2  &  0  & ${\pi^2\over 2}=4.93...$  & 2   & 4.42...  \cr
\noalign{\hrule}
}}$$
\noindent{\bf Table II:} Multi-river meander asymptotics. We have listed a few
values of $q=2\cos\pi f$, together with the corresponding values of 
$R(q,1)$ and $\alpha(q,1)$, and the numerical values of $R(q)$ obtained from
the direct enumeration results of Sect. 4.2 for comparison. Note that $R(q,1)\simeq R(q)+1/2$
with a good precision. 

\subsec{Exact Meander Asymptotics from Fully-Packed Loop Models coupled
to Two-dimensional Quantum Gravity}

In a recent work \ASY\ it has been noticed that the B$\&$W matrix model is the natural
random surface version of some Fully Packed loop model on the square lattice.
The latter is defined by assigning a color (B or W) to each edge of the square
lattice, in such a way that two edges of each color meet at each vertex. These
edges then form (Fully Packed) loops each of which is assigned a weight $p$ or $q$
for W and B loops respectively. The model is called the $FPL^2(p,q)$ model \JACO.
When defined on a random surface of genus zero, 
the model assigns colors to the edges of a random fatgraph with only vertices of 
the form $BWBW$ (crossing) or $BBWW$ (avoiding). The B$\&$W model of Sect. 6.1
does not have the second kind of vertices. Therefore the original Fully Packed 
loop model has been further restricted. 

The detailed study of the $FPL^2(p,q)$ model shows two remarkable facts: (i) it is
critical for all values of $0\leq p,q\leq 2$ (ii) it is represented in the
continuum limit by a Conformal Theory with central charge
\eqn\cfpl{ c_{FPL}(q,p)= 3-6\bigg( {e^2\over 1-e} + {f^2\over 1-f}\bigg)}
where $p=2\cos\pi e$ and $q=2\cos\pi f$. This was proved by mapping the
$FPL^2(p,q)$ model onto a three-dimensional height model, where the 
heights are defined in the center of each face, with an Amp\`ere-like
rule prescribing the transitions from one face to its neighbors. 
In the continuum limit,
the height variable becomes a three-dimensional free field (conformal theory
with central charge $c=3$), and the corrective weights assigning the factors
$p$ and $q$ per loop of each color account for the correction of $c$ by 
electric charges $e,f$.

The restriction we impose here on the $FPL^2(p,q)$ model on a random surface
amounts to restricting the height variable to only two dimensions instead of three.
The correct formula for the flat space Fully Packed Loop theory is therefore
\eqn\corecc{\eqalign{
c(q,p)&= 2-6\bigg( {e^2\over 1-e} + {f^2\over 1-f}\bigg)\cr
p&=2\cos\pi e\cr
q&=2\cos\pi f\cr}}
with $e,f\in [0,1/2]$ (i.e. $0\leq p,q \leq 2$).
We therefore state that the B$\&$W model of Sect. 6.1 is described in the planar
(large $N$) limit by the gravitational version of a conformal theory with central
charge \corecc, namely the same theory defined on fluctuating surfaces, that
have to be summed over statistically. Note that the $O(n)$ model, whose
gravitational version has been introduced in Sect. 2.3 is also described in the dense
phase by a conformal theory of central charge $c=1-6 g^2/(1-g)$ where $n=2\cos\pi g$.
So we are now dealing with a sort of double $O(n)$ model coupled to gravity.

The coupling to gravity of a conformal theory with central charge $c\leq 1$ has been
extensively studied within the context of non-critical string theory. 
The gravitational
theory has a new parameter $x$, called the cosmological constant, coupled to the area
of the surfaces we have to sum over. More precisely, the free energy for
a conformal theory coupled to gravity in genus zero reads
\eqn\gravi{ F={\rm Log}\, Z= \sum_{A\geq 0}x^A\sum_{{\rm connected}\
{\rm surfaces}\ \Gamma\ \atop {\rm of}\ 
{\rm area}\ A}Z_{CFT}(\Gamma)}
where $Z_{CFT}(\Gamma)$ denotes the partition function of the conformal
theory on the genus zero connected surface $\Gamma$. 
Comparing $Z$ with the B$\&$W model
partition function \bw, we see that $x$ plays the role of cosmological constant,
as $n=A$ are the areas of the tessellations dual to the fatgraphs of the model.
When the conformal theory has central charge $c$, the free energy \gravi\
has been shown to have a singularity of the form \critising\-\kpzone:
\eqn\singuF{ F\sim (x_c-x)^{2-\gamma} \qquad \gamma={1\over 12}(c-1-\sqrt{(1-c)(25-c)})}
when $x$ approaches some critical value $x_c$. This is easily translated into the 
large area asymptotics of the partition function of the model on surfaces of fixed area
\eqn\areaf{ F_A\sim {x_c^{-A} \over A^{3-\gamma}} }
Moreover, the operators of the conformal theory get ``dressed" by gravity, and
their correlation functions have singularities of the form 
\eqn\corre{ \langle \phi_{m_1} ...\phi_{m_k}\rangle \sim (x_c-x)^{\sum \Delta_{m_i}
-\gamma+2-k} }
where the ``dressed dimensions" $\Delta_m$ are related to the conformal dimensions $h_m$
of their undressed versions in the conformal theory through \KPZ\
\eqn\kpz{ \Delta_m={\sqrt{1-c+24 h_m} -\sqrt{1-c} \over \sqrt{25-c}-\sqrt{1-c}}}

We have now all the necessary material to compute the configuration exponents of 
all the meandric numbers of interest.
Applying the result \areaf\ to the central charge \corecc, we find the configuration
exponent of the multi-river meander polynomial \Ftotal\
\eqn\findi{\eqalign{
m_{2n}(q,p)&\sim {R(q,p)^{2n} \over n^{\alpha(q,p)}} \cr
\alpha(q,p)&=2+{1\over 12}\sqrt{1-c(q,p)}\big(\sqrt{25-c(q,p)}+\sqrt{1-c(q,p)}\big)\cr}}
as well as that of the multi-river semi-meander polynomial \corsem\
\eqn\semres{\eqalign{{\bar m}_{n}(q,p) &\sim 
{R(q,p)^{n} \over n^{{\bar\alpha}(q,p)} }\cr
{\bar \alpha}(q,p)&= \alpha(q,p)-1+2 \Delta_1 \cr}}
where $\Delta_1$ is the dressed dimension of the operator creating a white endpoint.
In the conformal theory, this operator is known to have the dimension \DUD\
\eqn\confod{ h_1= {1-e\over 16} -{e^2 \over 4(1-e)} }
where $p=2\cos\pi e$. We simply have to apply \kpz\ to \confod\ and substitute
the value of $\Delta_1$ back into \semres.

For $p=1$ ($e=1/3$) and $q$ arbitrary, we find
\eqn\exapone{\eqalign{
\alpha(q,1)&={2-f \over 1-f} \cr 
{\bar \alpha}(q,1)&={1\over 1-f} \cr}}
for all $q=2\cos\pi f$. The first line of \exapone\ agrees with the saddle point result
\randalph. The second line is readily obtained by noticing that $\Delta_1=0$ when
$e=1/3$, and therefore ${\bar \alpha}(q,1)=\alpha(q,1)-1$.

For $p=q=0$ ($e=f=1/2$), we get  the exact values of the meander and semi-meander
configuration exponents: 
\eqn\partitwo{ \eqalign{
\alpha(0,0)&= 2+{1\over 12}\sqrt{5}(\sqrt{5}+\sqrt{29}) \cr
{\bar \alpha}(0,0)&= 1+{1\over 24}\sqrt{11}(\sqrt{5}+\sqrt{29}) \cr}}

Note that the arguments of this section do not give any prediction for 
non-universal quantities such as $R(q,p)$ (which is expected to depend on
$q$ and $p$ explicitly, not just on $c(q,p)$).

Finally, for $p=0$ and $q$ arbitrary, we find:
\eqn\finoq{ \eqalign{
\alpha(q,0)&=2+{1\over 12}\sqrt{1-c(q)}\bigg(\sqrt{25-c(q)}+\sqrt{1-c(q)}\bigg) \cr
{\bar \alpha}(q,0)&=1+{1\over 24}\sqrt{3-4
c(q)}\bigg(\sqrt{25-c(q)}+\sqrt{1-c(q)}\bigg) \cr}}
with $c(q)$ given by \con. Note that the second line of \finoq\ breaks down when
$q=q_c$ corresponding to $c(q_c)=3/4$, namely 
\eqn\criq{ q_c=2 \cos\bigg(\pi {\sqrt{97}-1\over 48}\bigg) }
We identify this as the critical value of $q$ beyond which the winding
becomes relevant in semi-meanders, namely when circles (roads intersecting theriver
only once) dominate the semi-meander configurations.

\vfill\eject

\centerline{\bf PART C: Fluid Membrane Folding}
\vskip 1.truecm

We now turn to our second application of matrix integrals, having to do with 
the generation of foldable two-dimensional triangulations. 
These triangulations are a simple model for so-called tethered or fluid membranes,
objects of physical and biological interest.  

Although we now deal
with the folding of two-dimensional objects (as opposed to the one-dimensional
polymers of part B), we will rephrase the problem as that of enumerating 
vertex-tricolored triangulations. A suitable matrix model will be presented and solved 
using various techniques.

\newsec{Folding Triangulations}

As mentioned above, one usually distinguishes between two distinct models for
membranes:

\noindent{\bf Tethered Membranes:} are represented by reticulated networks, 
typically some domain of a regular lattice with physical vertices and rigid
bonds, allowed only to change their spatial configuration through folding.

\noindent{\bf Fluid Membranes:} are represented also by now irregular reticulated 
networks in which the valency of vertices is no longer fixed, this disorder 
accounting for the fluidity. But the bonds are still rigid, and the membrane
is still allowed to change its spatial configuration through folding.

In the following, we first briefly describe some results \DGTRI\ on the triangular
lattice folding
(tethered membrane folding), before turning to the problem of folding triangulations
\TRICO\ 
(fluid membranes). Here we only consider the
so-called ``phantom folding" of membranes,
i.e. we allow the network to interpenetrate itself, and are only interested in the
statistics of the final folded states, independently of the actual feasability
of the folding process. Moreover, the folding of the triangular lattice will be
two-dimensional, in that the folded configurations will be subsets of the original
lattice.

\subsec{Folding the Triangular Lattice}

\fig{A choice (a) for the tangent vectors of the triangular lattice, together
with the corresponding coloring of the edges by $1,2,3$.
This is the flat configuration of the membrane.
A folding configuration
(b) with the corresponding folded bonds (thick black lines) and edge coloring.
The three colors correspond to the three unit vectors
with vanishing sum represented above.}{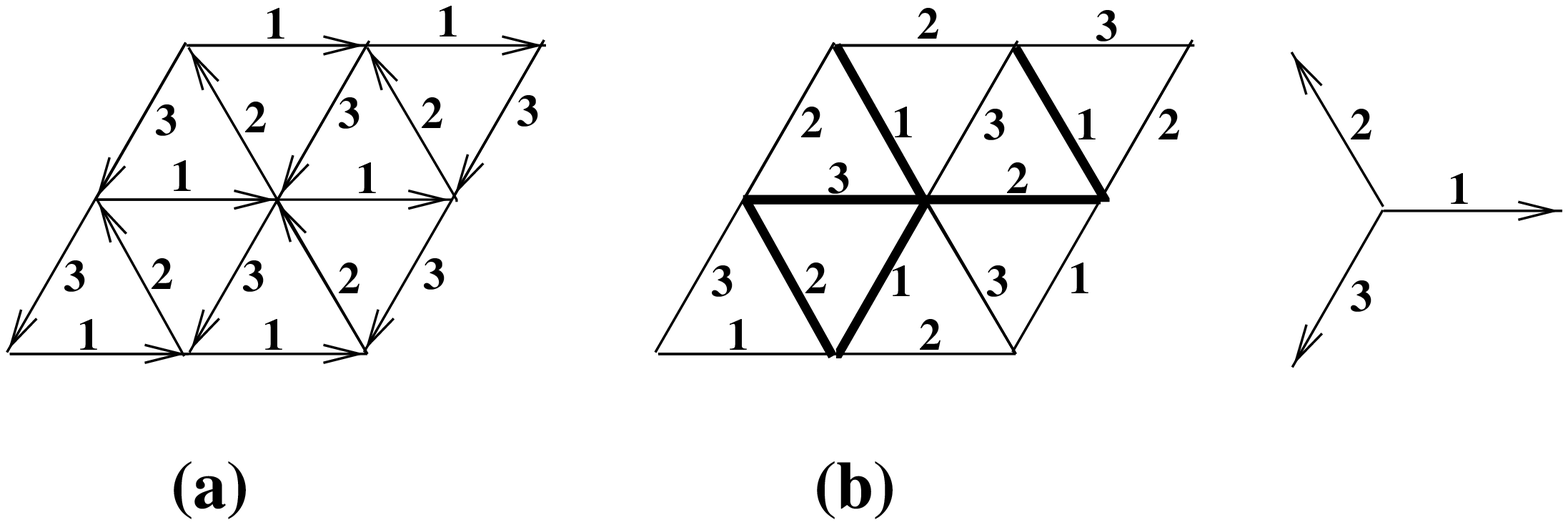}{10.cm}
\figlabel\tantri

We consider the triangular lattice, or rather a rhombus-shaped portion of it
with size $N\times P$.
Let us view this as a reticulated network,
with solid edges of unit length. Each such edge carries a unit tangent vector
$\vec{t}$, with the constraint that 
\eqn\rotang{\sum_{{\rm around}\ {\rm faces}} \vec{t}=\vec{0}} 
With the choice of tangent vectors indicated in Fig.\tantri\ (a),
a folding of the triangular lattice is a continuous map  $\rho:S\to \IR^2$,
preserving the length of the tangent vectors and
satisfying the condition \rotang\ around each triangular face of $S$.
Let $\vec{t}_1,\vec{t}_2,\vec{t}_3$ denote the unit tangent vectors
to a given face of $S$. Their images $\rho(\vec{t}_i)$ are three unit
vectors with vanishing sum, according to the face rule \rotang.
Fixing the image of one tangent vector of $S$ to be a given unit vector
$\vec{e_1}$, we see that the images of the tangent vectors of $S$
may only take three values $\vec{e_1},\vec{e_2},\vec{e_3}$, where
$\vec{e_1},\vec{e_2},\vec{e_3}$ are three unit vectors with vanishing sum,
hence forming angles of $2\pi/3$. Let us associate colors numbered $1,2,3$
to these three possible images.

A folding map $\rho$ of the triangular lattice is therefore a coloring of
its edges, with the three colors $1,2,3$, such that the three colors of edges
around each face are all distinct.
An example of such a coloring is given in Fig.\tantri\ (b) together with the
corresponding folding configuration.
The dual of this coloring model is the problem of tri-coloring the
edges of the hexagonal (honeycomb) lattice in such a way that the three
edges adjacent to each vertex are painted with distinct colors $1,2,3$.
It has been solved by Baxter \BAX, by use of the Bethe Ansatz.
Baxter's results yield in particular the exact value for the
thermodynamic entropy of folding per face of the triangular lattice
$s= \lim_{N,P\to \infty} {1\over NP} {\rm Log}\, Z_{N,P}$, where $Z_{N,P}$
denotes the total number of distinct folded configurations of our portion of 
lattice. It reads
\eqn\resbax{
s~=~{\rm Log}\bigg({\sqrt{3}\over 2\pi}
\Gamma(1/3)^{3/2}\bigg)}

This was originally proved by explicitly diagonalizing a large
(transfer) matrix, indexed by the coloring configurations of rows
of $N$ edges in the honeycomb lattice, and describing
the ``row-to-row transfer",
i.e. the allowed coloring configurations for two such neighboring rows.
The thermodynamic entropy \resbax\ is then the logarithm of
the largest (Perron-Frobenius) eigenvalue of this matrix.
The diagonalization is performed using a particular
ansatz for the form of the eigenvectors, the Bethe Ansatz.
The proof of \resbax\ being highly technical, we will not reproduce it
here, but refer the interested reader to the original papers \BAX.
Let us simply mention that this model is part of the class of
Two-dimensional Integrable Lattice Models, for which a Bethe
Ansatz solution exists.

A score of other lattice folding problems have been studied \SQDI. Actually, one can 
classify all the (compactly) foldable lattices \AMS, and define higher-dimensional
generalizations of their folding problems. Remarkably, all of them lead
to some equivalent coloring problems, also rephrased into Fully Packed
loop enumeration problems. 
For instance, in the case of folding the triangular lattice, we have seen that 
the model is equivalent to that of tricoloring the edges of the lattice. Concentrating
say on the colors $1$ and $2$, we see that edges of alternating colors $1212...$ form
loops on the dual (honeycomb) lattice, and moreover each vertex of the dual is visited
by exactly one of these loops: the loops $1212...$ are therefore fully packed, as
well as the $2323...$ and $1313...$ 
Remarkably, the same type of Fully Packed loop models
have emerged in our study of meanders (see Sect. 6.4 above).

\subsec{Foldable Triangulations}

Fluid membranes are modelled by irregular networks of
vertices linked by edges, in which the valencies of
the vertices are arbitrary, as well as the genus of the underlying
surface, which might have an arbitrary topology.
As advocated in Sect. 2,
matrix models provide us with a means of generating such graphs.

Let us restrict ourselves to networks with only triangular faces,
namely triangulations.
The initial question one can ask is: are all triangulations foldable? It is indeed
desirable to consider only triangulations with a large number of folded configurations,
otherwise the effect of folding might be wiped out in the limit of large size.
We therefore demand that our triangulations be {\it compactly} foldable, namely that 
one can fold them completely onto just one of their (equilateral triangular) faces.
With this constraint,
it is clear that not all triangulations turn out to be ``foldable".
Indeed, let us paint by three distinct colors $1,2,3$ the three vertices
of the image triangle, and paint accordingly the vertices of the preimages
under the folding map. This results in the tri-coloring of the vertices
of the initial triangulation, in such a way that the three colors around
each triangular face are distinct. So only the
vertex-tri-colorable triangulations
will be compactly foldable.
Another way of viewing this restriction is to recall that
we first need to attach tangent vectors
to the edges of the triangulation, in such a way that \rotang\ is satisfied.
It is straightforward to see that this is possible only if the vertices
of the triangulation are all even, as around such a vertex, we must have
an alternance of tangent vectors pointing to and from it.
This condition turns out to be sufficient in genus zero to grant the
tri-colorability of the triangulation. The situation in higher genus is unclear
\MIT.

Let us now introduce a generating function $Z(x_1,x_2,x_3;t;N)$ for
possibly disconnected vertex-tricolored triangulations of arbitrary genus, such that
\eqn\gentrico{\eqalign{ 
F(x_1,x_2,x_3;t;N)&={\rm Log}\, Z(x_1,x_2,x_3;t;N) \cr
&= \sum_{{\rm vertex}-{\rm tricolored}\atop
{\rm connected}\ {\rm triangulations}\ T} x_1^{n_1(T)} x_2^{n_2(T)}x_3^{n_3(T)}
{t^{A(T)\over 2}  N^{2-2h(T)}\over |{\rm Aut}(T)|} \cr}}
where $n_i(T)$ denote the total numbers of vertices of color $i$, $A(T)$
the total number of faces, $h(T)$ the genus and $|{\rm Aut}(T)|$ the
order of the symmetry group of the tricolored triangulation $T$. 

The construction of a matrix model to represent $Z(x_1,x_2,x_3;t;N)$ is
based on the following simple remark: in a given tri-colored triangulation $T$ 
if we remove say all the vertices of color $3$ and all edges connected to them,
we end up with a bi-colored graph, with unconstrained vertex valencies. 
Such bicolored graphs are easily built out of the Feynman graphs of a two 
Hermitian matrix model, say $M_1$ and $M_2$, the index $1$ and $2$ standing for the
color, with colored vertices
\eqn\verticolo{\eqalign{
{\rm Tr}(M_1^n)\ = \ \figbox{1.5cm}{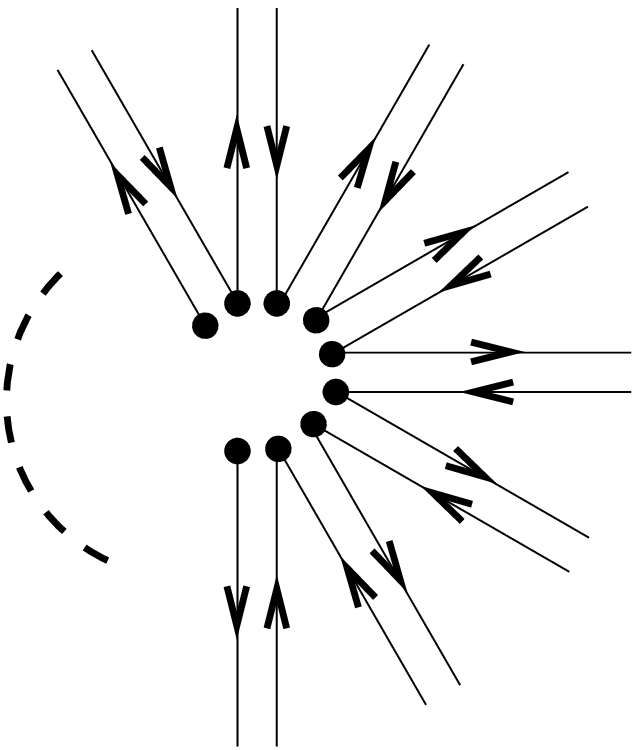}\ \  &\leftrightarrow N\, x_1 \cr
{\rm Tr}(M_2^n)\ = \ \figbox{1.5cm}{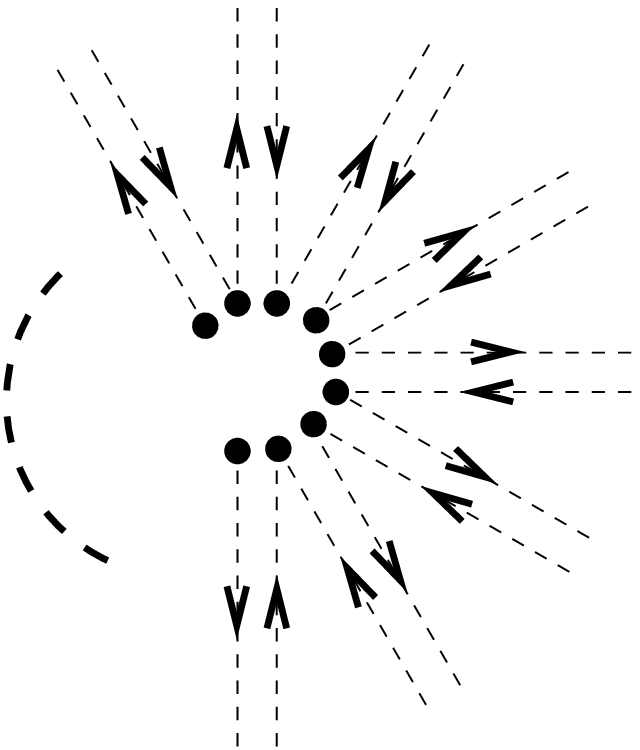}\ \  &\leftrightarrow N\, x_2 \cr}}
connected through propagators imposing the alternance of colors, namely
\eqn\propacol{ \langle (M_a)_{ij} (M_b)_{kl} \rangle\ 
=\ (1-\delta_{ab})\delta_{jk}\delta_{il} {t \over N} 
= \ \figbox{1.5cm}{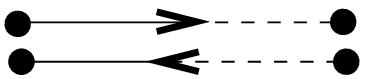}\ }
where $a,b=1,2$. Let us introduce the corresponding matrix integral, but 
keep $N$ fixed while the matrices are taken of size $n\times n$, $n$ possibly
different from $N$. This gives the partition function
\eqn\matrep{\eqalign{
Z_n(x_1,x_2;t;N)~&=~{1\over \varphi_n(t,N)}
\int dM_1 dM_2 e^{-N\, {\rm Tr}\, V(M_1,M_2;x_1,x_2,t)}\cr
V(M_1,M_2;x_1,x_2,t)~&=~x_1{\rm Log}(1-M_1)+x_2{\rm Log}(1-M_2) +{1\over t} M_1M_2 \cr}}
where the normalization factor $\varphi_n(t,N)$ ensures that $Z_n(0,0;t;N)=1$.
Clearly we must take the integral \matrep\ only at the level of formal series,
by expanding it in power series of $x_1$ and $x_2$, and computing the Gaussian integrals
with measure $dM_1dM_2 e^{-{N\over t}M_1M_2}$ by the Feynman procedure.
In particular, this gives the following rule for two-dimensional integrals
\eqn\twodco{\eqalign{ \langle x^\alpha y^\beta\rangle&=
{N\over t}\int dx dy e^{-{N\over t} xy} x^\alpha y^\beta \cr
&= \delta_{\alpha\beta} \Gamma(\alpha+1) \left({t\over N}\right)^{\alpha}\cr} }
The integral could be made rigorous and convergent by considering instead a normal
matrix $M_1\to M$, $M_2\to M^*$, but we will content ourselves with formal power
series anyway.

Let us now compute the Feynman graph expansion of the free energy
\eqn\fretrico{\eqalign{
F_n(x_1,x_2;t;N)~&=~{\rm Log}\, Z_n(x_1,x_2;t;N)\cr 
&=\sum_{{\rm bicolored}\ {connected}\atop
{\rm graphs}\ \Gamma} {1\over |{\rm Aut}(\Gamma)|} x_1^{n_1(\Gamma)} 
x_2^{n_2(\Gamma)} t^{E(\Gamma)} N^{(V(\Gamma)-E(\Gamma))} n^{F(\Gamma)}\cr
}}
where we have denoted by $n_i(\Gamma)$ the number of vertices of color $i$,
$V(\Gamma)=n_1(\Gamma)+n_2(\Gamma)$, $E(\Gamma)$ the number of edges, 
$F(\Gamma)$ the number of faces (the boundary of which are loops of running
indices, accounting for a factor $n$ each), and $|{\rm Aut}(\Gamma)|$ the
order of the symmetry group of the bicolored fatgraph $\Gamma$.
Adding a central vertex of color $3$ in the middle of each face of $\Gamma$,
and connecting it to all the vertices around the face with edges will result
in a vertex-tricolored triangulation $T$. The number of such added vertices
is nothing but $n_3(T)=F(\Gamma)$. Introducing
\eqn\link{ x_3={n\over N} }
we may rewrite
\eqn\rewfnco{ F_n(x_1,x_2;t;N) = F(x_1,x_2,x_3;t;N) }
by use of the Euler relation $2-2h(\gamma)=2-2h(T)=V(\Gamma)-E(\Gamma)+F(\Gamma)$
and the fact that $A(T)=2 E(\Gamma)$, as each edge of $\Gamma$ gives rise to two
triangles of $T$, one in each of the two faces adjacent to the edge. It is also
a simple exercise to show that $|{\rm Aut}(T)|=|{\rm Aut}(\Gamma)|$.
Considering \rewfnco\ as a formal power series of $t$ with polynomial coefficients
in $x_1,x_2,x_3$, we see that the knowledge of $F_n(x_1,x_2;t;N)$ for integer
values of $n$ determines completely the polynomial dependence on $x_3=n/N$
at each order in $t$. Hence computing $F_n(x_1,x_2;t;N)$ through
the integral formulation \matrep\ will yield the generating function for
compactly foldable triangulations.

\newsec{Exact Solution}

We now present the exact computation of the generating function $F(x_1,x_2,x_3;t;N)$
for compactly foldable triangulations. Although it can be obtained through 
orthogonal polynomial techniques generalizing that of Sect. 3.2, we choose
to present an alternative and powerful approach, using the Discrete Hirota equation,
in the form of a recursion relation for the quantities $F_n(x_1,x_2;t;N)$, in
terms of $n,Nx_1$ and $Nx_2$. We will then write some direct formal series expansion
for the solution, and use saddle-point techniques to extract the large $N$ behavior
of the free energy. This will give a nice formula for the generating function of 
vertex-tricolored triangulations of genus zero. 

\subsec{Discrete Hirota Equation}

The step zero in computing the integral \matrep\ is like in Sect. 3.1  
the reduction to eigenvalue integrals. Diagonalizing both matrices as
$M_i=U_im_iU_i^\dagger$, $m_i$ diagonal, $U_i$ unitary,
the change of variables $(M_1,M_2)\to (U_1,m_1;U_2,m_2)$ has the Jacobian
$J=\Delta(m_1)^2\Delta(m_2)^2$. We encounter a problem as the potential
$V(M_1,M_2;x_1,x_2,t)$ is not invariant under unitary conjugation, namely the
term ${\rm Tr}(M_1M_2)= {\rm Tr}(\Omega m_1\Omega^\dagger m_2)$, with the
unitary matrix $\Omega=U_2^\dagger U_1$.
So the integral over $U_i$ is no longer trivial, and yields a factor
\eqn\izint{ \int d\Omega e^{-{N\over t}{\rm Tr}(\Omega m_1\Omega^\dagger m_2))} 
\propto {\det(e^{-{N\over  t} m_{1,i} m_{2,j}})_{1\leq i,j \leq n}\over
\Delta(m_1)\Delta(m_2)} }
up to some normalization factor depending on $n$ and $N$ only.
Eqn.\izint\ is the celebrated Itzykson-Zuber formula of integration over the unitary
group \ITZ, itself a particular case of the Duistermaat-Heckmann localization formula \DH,
as the determinant in \izint\ may be viewed as a sum over the critical (saddle-)
points of the integrand, at $\Omega=P_\sigma$ the permutations of the eigenvalues.

The integral \matrep\ then reduces to
\eqn\leadz{ Z_n(x_1,x_2;t;N)~=~{1\over \psi_n(t,N)}
\int dm_1 dm_2 \Delta(m_1)\Delta(m_2)
e^{-N {\rm Tr}(V(m_1,m_2;x_1,x_2,t))} }
where the normalization factor $\psi_n(t,n)$ ensures that $Z_n(0,0;t;N)=1$.
It is easily derived by
expanding the two determinants as sums over permutations, with the result
\eqn\psideter{\eqalign{ \psi_n(t,n)~&=~\int dm_1dm_2 \Delta(m_1)\Delta(m_2)
e^{-Nu{\rm Tr}(m_1m_2)} \cr
&=~\sum_{\sigma,\tau \in S_n} {\rm sgn}(\sigma\tau)\prod_{i=1}^n
\int dm_{1,i}dm_{2,i}
m_{1,i}^{\sigma(i)-1} m_{2,i}^{\tau(i)-1} e^{-{N\over t} m_{1,i}m_{2,i}}\cr
&=~n! \prod_{i=1}^n {(i-1)!\over (N/t)^i} \cr}}
by use of \twodco.

Writing
\eqn\wriz{\eqalign{ Z_n&(x_1,x_2;t;N)~=~{1\over \psi_n(t,N)} \int
\Delta(m_1)\Delta(m_2) \cr
&\times \prod_{i=1}^n e^{-{N\over t} m_{1,i}m_{2,i}}
(1-m_{1,i})^{-a} (1-m_{2,i})^{-b}dm_{1,i} dm_{2,i}
\cr}}
with $a=Nx_1$, $b=Nx_2$,
and using the basic definition of determinants
\eqn\detergent{\eqalign{
\prod_{i=1}^n (1-m_{k,i})^{-a_k} \Delta(1-m_k)~&=~\det\, \bigg[
(1-m_{k,i})^{j-a_k-1} \bigg]_{1\leq i,j \leq n} \cr
&=~\sum_{\sigma \in S_n} {\rm sgn}(\sigma) \prod_{i=1}^n
(1-m_{k,i})^{\sigma(i)-a_k-1}\cr}}
for $k=1,2$, and the shorthand notation
\eqn\shorth{ Z_n(a,b)~=~ Z_n(x_1,x_2;t;N)\qquad a=Nx_1,b=Nx_2}
we finally get
\eqn\getzn{\eqalign{ Z_n(a,b)~&=~{1\over \psi_n(t,N)}
\sum_{\sigma,\tau\in S_n} {\rm sgn}(\sigma\tau)
\prod_{i=1}^n \int dm_{1,i} dm_{2,i} \cr
&(1-m_{1,i})^{\sigma(i)-a-1} (1-m_{2,i})^{\tau(i)-b-1}
e^{-{N\over t}m_{1,i} m_{2,i}}\cr
&=~{n!\over \psi_n(t,N)}\sum_{\nu\in S_n}{\rm sgn}(\nu)\cr
&\prod_{i=1}^n
\int dx dy (1-x)^{i-a-1} (1-y)^{\nu(i)-b-1}e^{-{N\over t}xy} \cr}}
where we have set $\nu=\tau \sigma^{-1}$, with the same signature
as $\sigma\tau$, and explicitly factored out the sum over $\sigma$.
Moreover, the dummy integration variables have been rebaptized
$x$ and $y$, and the integral can be computed by expanding the integrand
as a power series of $x,y$ and then using term by term the prescription \twodco.
The partition function takes therefore the form
\eqn\deterfor{ Z_n(a,b)~=~{n!\over \psi_n(t,N)}\, D_n(a,b) }
where $D_n(a,b)$ is the $n\times n$ determinant
\eqn\Ddefi{
D_n(a,b)~=~\det\bigg[ \int dx dy (1-x)^{i-a-1}
(1-y)^{j-b-1} e^{-Nuxy} \bigg]_{1\leq i,j \leq n} }
and $\psi_n(t,N)/n!=D_n(0,0)=\prod_{1\leq i\leq n} (i-1)!/(N/t)^i$.

The Hirota equation \WIG\ is simply the rephrasing in terms of $D_n(a,b)$
of the following identity satisfied by any $(n+1)\times (n+1)$ determinant $D$
and the minors $D_{i,j}$ obtained by erasing the $i$-th row and $j$-th column,
as well as the minors $D_{i_1,i_2;j_1,j_2}$ obtained by erasing the rows $i_1,i_2$
and columns $j_1,j_2$ in $D$:
\eqn\qadradet{ D\, D_{1,n+1;1,n+1} ~=~
D_{n+1,n+1}\, D_{1,1} - D_{1,n+1}\, D_{n+1,1}}
When expressed in terms of $D=D_{n+1}(a+1,b+1)$, this implies the
quadratic equation
\eqn\hiro{\eqalign{
D_{n+1}&(a+1,b+1)D_{n-1}(a,b)\cr 
&=D_n(a+1,b+1)D_n(a,b)-
D_n(a,b+1)D_n(a+1,b)\cr} }
Finally, using $\psi_{n+1}(t,N)\psi_{n-1}(t,N)/\psi_n(t,N)^2=n/(Nu)=nt/N$ from
\psideter, we get the Hirota Bilinear equation for $Z_n(a,b)$ by substituting
\deterfor\ into \hiro:
\eqn\hirota{
n {t\over N}\, Z_{n+1}(a+1,b+1)\, Z_{n-1}(a,b)
=Z_n(a+1,b+1)\, Z_n(a,b)-Z_n(a,b+1)\, Z_n(a+1,b)}
This recursion relation determines the partition function $Z_n(a,b)$ entirely,
once we apply the following initial conditions. Let 
\eqn\nergy{F_n(a,b)={\rm Log}\, Z_n(a,b)=\sum_{m\geq 1}
(t/N)^m \omega_m(a,b,n) }
be the generating function for connected tricolored triangulations, with
$a=Nx_1,b=Nx_2$ and $n=Nx_3$. 
Then, introducing the shorthand notation $\delta_xf(x)=f(x+1)-f(x)$ for finite
differences, the Hirota equation \hirota\ turns into a finite difference equation for
$F_n(a,b)$:
\eqn\finidif{ \delta_{a}\delta_{b} 
F_n(a,b)~=~-{\rm Log}\big(1-n {t\over  N}
e^{\delta_n F_n(a+1,b+1) -\delta_n F_{n-1}(a,b)}
\big) }
This turns into a nonlinear recursion relation for the coefficients $\omega_m(a,b,n)$.
But thanks to their interpretation in terms of tricolored triangulation counting,
namely $\sum \omega_m(Nx_1,Nx_2,Nx_3)(t/N)^m=\sum N^{2-2h} x_1^{n_1} x_2^{n_2} x_3^{n_3} 
\#($tricol.triang.$)$,  
all of these are polynomials of $a,b,n$, and all have at least a term $abn$ in
factor, as there is always at least one vertex of each color $1,2,3$ in such a
triangulation. This allows for performing the discrete integration step involved
in the recursion, and yields the complete solution for $F_n(a,b)$ as a formal
power series of $t$. The first few terms read:
\eqn\firefew{\eqalign{
\omega_{1}~&=~na b  \cr
\omega_{2}~&=~{nab \over 2}(n+a+b) \cr
\omega_{3}~&=~{n a b\over 3}(n^2+3(a+b)n+a^2+3 ab +b^2+1)\cr
\omega_{4}~&=~
{n ab \over 4}(n^3+6(a+b)n^2+(6a^2+17 ab +6 b^2+5)n\cr
&+(a+b)(a^2+5 ab+b^2+5))\cr
}}
Note the symmetry in $a,b,n$, now manifest.
In the limit of large $N$, the genus zero free energy 
\eqn\gezerfre{f_0(x_1,x_2,x_3;t)=\lim_{N\to \infty} {1\over N^2} F_{Nx_3}(Nx_1,Nx_2)}
satisfies the differential equation
\eqn\difeqef{\partial_{x_1}\partial_{x_2} f_0~=~ -{\rm Log}\big( 1-
t x_3e^{\partial_{x_3}(\partial_{x_1}+
\partial_{x_2}+\partial_{x_3})f_0}\big) }
as a consequence of \finidif.

\subsec{Direct Expansion and Large N Asymptotics}

Let us explicitly compute the partition function $Z_n(a,b)$ \deterfor\
by expressing the determinant \Ddefi\ as a formal series expansion in $t$.
In the integrand of \Ddefi, we expand all terms of the form
\eqn\forterm{ (1-u)^{-c}=\sum_{k\geq 0} u^k {\Gamma(k+c)\over \Gamma(c) k!} }
for $(u,c)=(x,a+1-i)$ and $(y,b+1-j)$ respectively, and then compute the defining
integral by use of the prescription \twodco. After a little algebra, we arrive at
\eqn\redet{\eqalign{
&D_n(a,b)~=~\det\bigg[\sum_{k\geq 0}{1\over k!}
{\Gamma(k+a-i+1)\over \Gamma(1+a-i)}
{\Gamma(k+b-j+1)\over \Gamma(1+b-j)} \big( {t\over N}\big)^{k+1}
\bigg]_{1\leq i,j\leq n} \cr
&=\sum_{k_1,...,k_n\geq 0} \prod_{i=1}^n
{(t/N)^{k_i+1}\Gamma(k_i+a-i+1)\over k_i!\, 
\Gamma(1+a-i)\Gamma(1+b-i)}
\det\big[\Gamma(k_i+b-j+1)
\big]_{1\leq i,j \leq n}\cr}}
Factoring $\Gamma(k_i+b-n+1)$ out of each line of the remaining determinant, we 
are left with the computation of the determinant
\eqn\deterleftz{
\det \big[(k_i+b-j)(k_i+b-j-1)...(k_i+b-n+1)\big]~=~
\det \big[ q_{n-j}(k_i) \big] }
where the polynomials $q_m$ are monic of degree $m$, and as such satisfy
$\det [ q_{n-j}(k_i) ]=\det[ k_i^{j-1}]=\Delta(k)$ from \vandereq, 
the Vandermonde determinant of the $k$'s.
Repeating the above trick with the columns of the determinant, we finally get the 
symmetric expression
\eqn\zndetf{
Z_n(a,b)~=~
\sum_{k_1,...,k_n\geq 0} \Delta(k)^2 \prod_{i=1}^n
{(t/N)^{k_i+1-i}\over i!\ k_i!}
{\Gamma(k_i+a-n+1)\over \Gamma(1+a-i)}{\Gamma(k_i+b-n+1)\over \Gamma(1+b-i)}}

By construction, this formal series of $t$ is a solution to the Hirota equation
\hirota. Note the remarkable similarity between the expansion \zndetf\ and the
reduction to eigenvalues of the one-matrix integral \obtz, except that the 
integration over eigenvalues is replaced by a sum over non-negative integers.
We indeed have the same two types of terms: (i) the squared Vandermonde determinant,
i.e. the repulsion term,
and (ii) the product of ratios of gamma functions, that can be exponentiated into 
a potential term.
The expansion \zndetf\ allows us for a computation of the large $N$ limit of 
the free energy, by applying the saddle-point techniques. Indeed, writing
$k_i=N\alpha_i$, we may estimate the expansion by a real integral over
the $\alpha$'s when $N$ is large, of the form
$\int d\alpha_1...d\alpha_n \exp(-N^2 S(\{\alpha_i\}))$, where the index $i$ itself
ranging from $1$ to $n=Nx_3$ may be written as $i=Ns$, $s\in [0,1]$. Expressing that this
integral is dominated by the minimum of $S$, it is a straightforward, though tedious
calculation to get the large $N$ asymptotics of $Z$. We refer the reader to ref. \TRICO\ 
for details, and simply give the beautifully simple result here. The genus
zero free energy of the model \gezerfre\ satisfies the following
\eqn\fgenzer{\eqalign{
t(t\partial_t)^2 f_0(x_1,x_2,x_3;t)~&=~ 
F_1 F_2 F_3\cr 
F_1(1-F_2-F_3)~&=~ t x_1 \cr
F_2(1-F_3-F_1)~&=~ t x_2 \cr
F_3(1-F_1-F_2)~&=~ t x_3 \cr}}
where $F_i\equiv F_i(x_1,x_2,x_3;t)=t x_i+ O(t^2)$ are formal series of $t$
with polynomial coefficients of the $x$'s. Actually it is easy to see that 
$F_i$ is the generating function for tricolored rooted trees, whose
root has color $i$. This result still awaits a good combinatorial
interpretation.
One way of proving the result \fgenzer\ {\it a posteriori} is to check that it
satisfies the differential equation \difeqef.
As an independent check, in the particular case $x_1=x_2=x_3=z$ we find
$F_1=F_2=F_3=F$ and $F(1-2F)=tz$, in agreement with a former result of
Tutte \TUT. 

Analyzing the critical properties of $f_0$ as a function of $t$, we find a
singular behavior of the form
$f_0(t)\sim (t_c-t)^{5/2}$ for generic values of the $x$'s 
($x_i>0$ and $x_i\neq x_j$ for $i\neq j$). This leads to a number of
tricolored triangulations behaving as $T(x_1,x_2,x_3)^A/A^{7/2}$ in terms of
their fixed area $A$.
We recover therefore the value $\gamma=-1/2$ \kpzone\ of the pure gravity,
meaning that the tricoloring constraint does not affect the configuration exponent
of triangulations, but does affect the leading behavior through the
function $T(x_1,x_2,x_3)\neq 12$.

\newsec{Conclusion}

The combinatorial applications of matrix integrals are various and many, and it would be
impossible to even enumerate them all here. In these notes we have chosen to concentrate
on folding problems with simple physical interpretations, but this is mainly a matter of
taste. 

In the study of meanders, we have shown that even an almost one-dimensional 
problem had to be first viewed as a random graph (or surface) problem, and then 
matrix and quantum gravity techniques have allowed for obtaining many interesting results.

In the case of foldable triangulation enumeration, the matrix model has allowed
for the derivation of a very simple formula for the genus zero counting function,
that still has to be interpreted combinatorially. 

The main lesson we would like to draw is that the matrix models give a maybe less
intuitive but definitely different angle on graph-related combinatorial problems. 

\vskip 1.truecm
\noindent{\bf Acknowledgements:} I would like to thank A. Its and P. Bleher, organizers of 
the semester ``Random Matrices and Applications" at M.S.R.I. (Spring 1999) for their
hospitality and their suggestion to write these notes. This work was partially supported by
the NSF grant PHY-9722060. 

\listrefs
\bye